\documentclass[aps,prd,preprintnumbers,twocolumn,showpacs,nofootinbib]{revtex4}
\usepackage{amsmath,amssymb}
\usepackage{mathrsfs}
\usepackage[dvips]{graphicx}
\usepackage{enumerate}
\usepackage{bm}

\begin{document}

\preprint{CHIBA-EP-184v2, July 2010}
\title{Toward a first-principle derivation of 
confinement and chiral-symmetry-breaking crossover transitions
 in QCD
}

\author{Kei-Ichi Kondo$^{1,2}$} 
\email{kondok@faculty.chiba-u.jp}
\affiliation{$^1$Department of Physics,  Chiba University, Chiba 263-8522, Japan}
\affiliation{$^2$Department of Physics, The University of Tokyo, Tokyo 113-0033, Japan}

\date{\today}

\begin{abstract}

We give a theoretical framework to obtain a low-energy effective theory of quantum chromodynamics (QCD) towards a first-principle derivation of confinement/deconfinement and chiral-symmetry breaking/restoration crossover transitions. 
In fact, we demonstrate that an effective theory obtained using simple but non-trivial approximations within this framework enables us to treat both transitions simultaneously on equal footing. 
A resulting effective theory is regarded as a modified and improved version of nonlocal Polyakov-loop extended Nambu-Jona-Lasinio (nonlocal PNJL) models proposed recently by Hell,  R\"ossner, Cristoforetti and Weise, and Sasaki, Friman and Redlich, extending the original (local) PNJL model by Fukushima and others. 
A novel feature is that the nonlocal NJL coupling depends explicitly on the temperature and Polyakov loop, which affects the entanglement between confinement and chiral symmetry breaking,  together with the cross term introduced through the covariant derivative in the quark sector considered in the conventional PNJL model.  The chiral symmetry breaking/restoration transition is  controlled by the nonlocal NJL interaction, while the confinement/deconfinement transition in the pure gluon sector is specified by the nonperturbative effective potential for the Polyakov loop obtained recently by  Braun, Gies, Marhauser and  Pawlowski. 
The basic ingredients are a reformulation of QCD based on new variables and the flow equation of the Wetterich type in the Wilsonian renormalization group. 
This framework can be applied to investigate the QCD phase diagram at finite temperature and density. 
\end{abstract}


\maketitle

\section{Introduction}\label{sec:intro}

The relation between confinement and chiral symmetry breaking is one of the long-standing puzzles in theoretical physics. 
Recently,  strong interest on this issue revived in extreme environments especially at high temperatures and baryon densities,   
stimulated by the heavy-ion programs at GSI, CERN SPS, RHIC and  LHC, see e.g., \cite{KS04,YHM05} for a review.
Quantum chromodynamics (QCD) for strong interactions is a fundamental theory for solving this problem.

In pure Yang-Mills theory, i.e., in the limit of infinitely heavy quark mass $m_q \rightarrow \infty$ of QCD, the Polyakov loop average $\langle L \rangle$, i.e., the vacuum expectation value of the Polyakov loop  operator $L$, can be used as a criterion for quark confinement \cite{Polyakov78}. 
The Polyakov loop operator $L$ is a gauge invariant operator charged under the center group $Z(N_c)$ of the color gauge group $SU(N_c)$.  The Polyakov loop average $\langle L \rangle$ vanishes $\langle L \rangle=0$ and quarks are confined at low temperatures $T<T_d$ where the global center symmetry $Z(N_c)$ is intact, while it is nonzero $\langle L \rangle \not= 0$ and quarks are deconfined at high temperature $T>T_d$ where the global center symmetry $Z(N_c)$ is spontaneously broken. 
Thus, we can define $T_d$ as a critical temperature for confinement/deconfinement phase transition. 

When dynamical quarks in the fundamental representation of the gauge group are added to the Yang-Mills theory, the center symmetry is no longer exact. 
On the other hand, QCD with massless quarks $m_q \rightarrow 0$ exhibits chiral symmetry $SU(N_f)_L \times SU(N_f)_R$. 
The chiral condensate $\langle \bar\psi \psi \rangle$, i.e., the vacuum expectation value of a gauge-invariant composite operator $\bar\psi \psi$, is used as an order parameter for chiral symmetry breaking.
The chiral condensate $\langle \bar\psi \psi \rangle$ is nonzero $\langle \bar\psi \psi \rangle \not= 0$ at low temperatures $T<T_\chi$ where the chiral symmetry is spontaneously broken, while it vanishes $\langle \bar\psi \psi \rangle = 0$ at high temperature $T>T_\chi$ where the chiral symmetry is restored.  
Thus, we can define $T_\chi$ as a critical temperature for chiral-symmetry breaking/restoration phase transition. 

For realistic quark mass (with finite and nonzero $m_q$: $0 < m_q <\infty$), there are no exact symmetries directly related to the phase transitions, since both the center and chiral symmetries are explicitly broken, and $\left< L \right>$ and $\left< \bar\psi \psi \right>$ are approximate order parameters. 
In this case, there is no critical temperature  $T_c$ in strict sense, and the transition  can be a crossover transition for which the pseudo critical temperature $T_c^*$ is defined such that the susceptibility takes the maximal value at $T=T_c^*$.
If quarks are in the fundamental representation, deconfinement (a rise in the Polyakov loop average) happens at the temperature where the chiral symmetry is restored (chiral condensate decreases rapidly). 
The chiral and deconfinement transitions seems to coincide, 
$T_d^*=T_\chi^* \simeq T_c$ \cite{Karsch02,KL94},  
although the property of the phase transition, e.g., the critical temperature and the order of the transition depend on  numbers of color $N_c$ and flavor $N_f$. 
Whereas, for quarks in the adjoint representation, deconfinement and chiral-symmetry restoration do not happen at the same temperature, rather $T_\chi^* \gg  T_d^*$ \cite{KL98}. 
Although there exist theoretical considerations on the interplay between chiral symmetry breaking and confinement at zero baryon density \cite{Casher79}, the underlying reasons for the coincidence are still unknown and uncertain at nonzero baryon density \cite{KST99,MP07,HKS10}. 

The hadronic properties, especially, chiral dynamics at low energy have been successfully described by chiral effective models such as the linear sigma model \cite{Lee72}, the Nambu-Jona-Lasinio (NJL) model \cite{NJL61,Klevansky92,HK94}, the chiral random matrix model \cite{SV93}, chiral perturbation theory \cite{GL84} and so on. 
However, those models based on chiral symmetry lack any dynamics coming from confinement dictated by the Polyakov loop, although there are some efforts to clarify the interplay between chiral dynamics and the Polyakov loop 
\cite{MO95,MST03}.

Recent chiral effective models with the Polyakov loop degrees of freedom augmented called the Polyakov loop--extended NJL  (PNJL) model or  quark-meson (PQM) model  \cite{Fukushima04,MRAS06,RTW05,SFR06,KKMY07,HRCW08,BBRV07,SPW07} are successful  from a phenomenological  point of view to incorporate a coupling between the chiral condensate and the Polyakov loop. 
However, these PNJL/PQM models are still far from treating the chiral condensate and the Polyakov loop on an equal footing, except for the work \cite{SPW07} where the backcoupling of the matter sector to the glue sector was discussed by changing the phase transition parameter.
In fact, the gluonic part in these models has several fitting parameters which are determined only from lattice QCD data.

Here we must mention a preceding work for a first-principle derivation of confinement/deconfinement and chiral-symmetry-breaking/restoration crossover phase transition based on the flow equation \cite{Wetterich93}  of the functional renormalization group \cite{BTW00,Gies06} given by Braun, Haas, Marhauser and Pawlowski \cite{BHMP09} for the full dynamical QCD with 2 massless flavors (at zero and imaginary chemical potential).  In this work, the Yang-Mills theory is fully coupled to the matter sector by taking into account the Polyakov-loop effective potential  \cite{Weiss81} obtained in a nonperturbative way put forward by Braun, Gies and Pawlowski \cite{BGP07} and Marhauser and Pawlowski \cite{MP08}.

The main purpose of this paper is to provide a theoretical framework (a reformulation of QCD) which enables one to describe in a unified way the chiral dynamics and  confinement signaled by the Polyakov loop.
We give  an important step towards a first-principle derivation of confinement/deconfinement and chiral-symmetry breaking/restoration crossover transition. 
In fact, we demonstrate that a low-energy effective theory of QCD obtained in  simple  but non-trivial approximations within this framework enables one to treat both transitions simultaneously on equal footing.

The basic ingredients in this paper are a reformulation of QCD based on new variables \cite{Cho80,DG79,FN99,Shabanov99,KMS06,KMS05,Kondo06,KSM08,KKMSSI05,IKKMSS06,SKKMSI07,SKKMSI07b,KSSMKI08,SKS09,Cho00,Kondo08,KS08,Kondo08b} and the flow equation of the Wetterich type in the Wilsonian renormalization group \cite{Wetterich93,BTW00,Gies06}. 
The reformulation was used to confirm quark confinement in pure Yang-Mills theory at zero temperature and zero density  based on a dual superconductor picture \cite{dualsuper}. 
In this paper, it is extended to QCD at finite temperature and density. 
In principle, our framework can be applied to any color gauge group and arbitrary number of flavors. For technical reasons, however, we study two color QCD with two flavors in this paper.  The three color and/or three flavor case will be studied in a subsequent paper. 
In future publications, this framework will be applied to investigate the QCD phase diagram at finite density. 
We hope that this paper will give an insight into this issue complementary to other works, e.g.,  \cite{BHMP09}.

In sec. II, we give a reformulation of QCD written in terms of new variables and explain why the reformulated QCD is efficient to study the interplay between confinement and chiral-symmetry  breaking. 

In sec. III, we choose a specific gauge (modified Polyakov gauge) to simplify the representation of the Polyakov loop.
We can choose any gauge to calculate the Polyakov loop average and the chiral condensate, since both are gauge-invariant quantities and should not depend on the gauge chosen.  

In sec. IV, we give a definition of the Polyakov loop operator and examine how the Polyakov loop average is related to the average of the time-component of the gauge field. 

In sec.~V and sec.~VI, we study the confinement/deconfinement phase transition in pure $SU(2)$ Yang-Mills theory at finite temperature.
We exploit the Wilsonian renormalization group in our framework to obtain the effective potential $V_{\rm eff}$ of the Polyakov loop $L$, whose minimum gives the Polyakov loop average $\langle L \rangle$. 
It is known that the Weiss potential $V_{\rm W}$ \cite{Weiss81} calculated in the perturbation theory to one loop exhibits spontaneous center-symmetry breaking, i.e., deconfinement, irrespective of the temperature $T$. 
This result can be used at high temperature where the perturbation theory will be trustworthy due to asymptotic freedom, while nonperturbative approach is necessary to treat the low-temperature case. 
The Weiss potential can be improved according to the Wilsonian renormalization group to obtain a nonperturbative effective potential which is valid even at low temperature.

In sec. V, we write down the flow equation of the Wetterich type for the effective potential of the Polyakov loop in our framework. 
In fact, the effective potential obtained by solving the flow equation in a numerical way  shows the existence of confinement phase below a certain temperature $T_d$. 
This solution was shown for the first time by Marhauser and Pawlowski \cite{MP08} and by Braun, Gies and  Pawlowski \cite{BGP07}, see  \cite{MO97} for the previous works.  
In this sense, this section is nothing but the translation of their  results \cite{MP08,BGP07} into our framework.

In sec. VI, we give a qualitative understanding for the confinement/deconfinement transition given in sec. V based on the Landau-Ginzburg argument.  We answer a question why  the center-symmetry restoration occurs as the temperature is decreased,  by observing the flow equation for the coefficient of the effective potential.  

In sec. VII, we describe the low-energy effective interaction among quarks  by a nonlocal version of the (gauged) NJL model in which  the effect of confinement is  explicitly incorporated through the Polyakov loop dependent nonlocal interaction.   
The resulting effective theory is regarded as a modified and improved version of  nonlocal Polyakov-loop extended Nambu-Jona-Lasinio (PNJL) models proposed recently by Hell,  R\"ossner, Cristoforetti and Weise \cite{HRCW08}, Sasaki, Friman and  Redlich \cite{SFR06}, and Blaschke, Buballa, Radzhabov and Volkov \cite{BBRV07}, extending the original (local) PNJL model by Fukushima \cite{Fukushima04}. 
The nonlocal (gauged) NJL model can be converted to the nonlocal (gauged) Yukawa model to be bosonized to study the chiral dynamics.

In sec. VIII, we show that the nonlocal NJL interaction among quarks  becomes temperature dependent through the coupling to the Polyakov loop. 
This is a first nontrivial indication for the entanglement between the chiral symmetry breaking and confinement. 
This feature was overlooked in  conventional PNJL models. 

In sec. IX, we consider how to understand the entanglement between confinement and chiral symmetry breaking in our framework. 
This is just a short sketch for our strategy following the line given in the preceding sections. 

The final section is used to summarize the results and give some perspective in the future works. 
Some technical materials are collected in Appendices.

\section{Reformulation of QCD}\label{sec:reformulation}

To fix the notation, we write the action of QCD in terms of the gluon field $\mathscr{A}_\mu$ and the quark field $\psi$:
\begin{align}
\label{eq:QCDaction}
 S_{\rm QCD} =& S_{\rm q}  + S_{\rm YM} ,
\nonumber \\
 S_{\rm q}  :=& \int d^Dx \bar{\psi} (i\gamma^\mu \mathcal{D}_\mu[\mathscr{A}] -\hat{m}_q +  \mu_q \gamma^0) \psi ,
\nonumber\\
 S_{\rm YM} :=&  \int d^Dx  \frac{-1}{2} {\rm tr}(\mathscr{F}_{\mu\nu}[\mathscr{A}] \mathscr{F}^{\mu\nu}[\mathscr{A}]) 
 ,
\end{align}
where  $\psi$ is the quark field, $\mathscr{A}_\mu=\mathscr{A}_\mu^A T_A$ is the gluon field with $su(N_c)$ generators $T_A$ for the gauge group $G=SU(N_c)$ ($A=1, \cdots, {\rm dim}SU(N_c)=N_c^2-1$), $\hat{m}_q$ is the quark mass matrix,  $\mu_q$ is the quark chemical potential, $\gamma^\mu$ are the Dirac gamma matrices ($\mu=0, \cdots, D-1$), $\mathcal{D}_\mu[\mathscr{A}]:=\partial_\mu - ig \mathscr{A}_\mu$ is the covariant derivative in the fundamental representation, $\mathscr{F}_{\mu\nu}[\mathscr{A}]:=\partial_\mu \mathscr{A}_\nu - \partial_\nu \mathscr{A}_\mu -i g [\mathscr{A}_\mu ,  \mathscr{A}_\nu]$ is the field strength and $g$ is the QCD coupling constant.  
In what follows, we suppress the spinor, color and flavor indices. 

The main purpose of this paper is to give a theoretical framework for extracting a low-energy effective theory which enables one to discuss the confinement/deconfinement and chiral-symmetry breaking/restoration (crossover) transition simultaneously on an equal footing.  
We reformulate QCD in terms of new variables which are efficient for this purpose.  
We start with decomposing the original $SU(N)$ Yang-Mills field $\mathscr{A}_\mu(x) =\mathscr{A}_\mu^A(x)  T_A$ into two pieces $\mathscr{V}_\mu=\mathscr{V}_\mu^A(x)  T_A$ and $\mathscr{X}_\mu=\mathscr{X}_\mu^A(x)  T_A$:
\begin{equation}
  \mathscr{A}_\mu(x) = \mathscr{V}_\mu(x) + \mathscr{X}_\mu(x)  ,
  \label{decomp}
\end{equation}  
to rewrite the original QCD action into a new form:
\begin{align}
 S_{\rm q}   =& \int d^Dx \Big\{ \bar{\psi} (i\gamma^\mu \mathcal{D}_\mu[\mathscr{V}] -\hat{m}_q +  \mu_q \gamma^0) \psi  
 + g\mathscr{J}^{\mu} \cdot \mathscr{X}_\mu \Big\} ,
\nonumber\\
 S_{\rm YM}  =& \int d^Dx  \Big\{ \frac{-1}{4} (\mathscr{F}_{\mu\nu}^A[\mathscr{V}])^2
-     \frac{1}{2} \mathscr{X}^{\mu A}  Q_{\mu\nu}^{AB} \mathscr{X}^{\nu B} 
\nonumber\\&
-  \frac14 (i g [ \mathscr{X}_\mu , \mathscr{X}_\nu ])^2 \Big\}
 + S_{\rm FP},
\label{QCDaction2}
\end{align}
where 
$\mathscr{J}^{\mu A}  := g\bar{\psi} \gamma^\mu T_A  \psi $ is the color current, 
$D_\mu[\mathscr{V}] := \partial_\mu - ig[\mathscr{V}_\mu , \cdot ]$ is the covariant derivative in the adjoint representation and 
\begin{align} 
Q_{\mu\nu}^{AB}[\mathscr{V}]  :=& G^{AB} [\mathscr{V}] g_{\mu\nu} 
+ 2gf^{ABC} \mathscr{F}_{\mu\nu}^{C}[\mathscr{V}] , 
\nonumber\\
G^{AB}[\mathscr{V}]  :=& - (D_\rho[\mathscr{V}]D^\rho[\mathscr{V}])^{AB}   
\nonumber\\
=& -(\partial_\rho \delta^{AC}+gf^{AEC}\mathscr{V}_\rho^E) (\partial^\rho \delta^{CB}+gf^{CFB}\mathscr{V}^{\rho F})   
\nonumber\\
=& - \partial_\rho^2 \delta^{AB} + g^2 f^{AEC}f^{BFC} \mathscr{V}_\rho^E \mathscr{V}^{\rho F} 
\nonumber\\
&+ 2g f^{ABE} \mathscr{V}_\rho^E \partial^\rho + gf^{ABE} \partial^\rho \mathscr{V}_\rho^E 
 .
\label{Q}
\end{align}
In what follows we use the notation $\mathscr{A} \cdot \mathscr{B}$ for two Lie-algebra valued functions $\mathscr{A}=\mathscr{A}^A T_A$ and $\mathscr{B}=\mathscr{B}^A T_A$  in the sense that 
$\mathscr{A} \cdot \mathscr{B} := \mathscr{A}^A  \mathscr{B}^A  = 2 {\rm tr}(\mathscr{A} \mathscr{B})$
and especially $\mathscr{A}^2:=\mathscr{A} \cdot \mathscr{A}= \mathscr{A}^A  \mathscr{A}^A$.

Historically, the decomposition of Yang-Mills theory into new variables has been proposed by Cho \cite{Cho80} and Duan and Ge \cite{DG79} independently, and readdressed later by Faddeev and Niemi \cite{FN99}.  
The decomposition was further developed by Shabanov \cite{Shabanov99}.

The decomposition (\ref{decomp}) is performed such that  
$\mathscr{V}_\mu$ transforms under the gauge transformation just like the original gauge field $\mathscr{A}_\mu$: 
\begin{equation}
  \mathscr{V}_\mu(x)   \rightarrow \mathscr{V}_\mu^\prime(x) = \Omega(x) (\mathscr{V}_\mu(x) + ig^{-1} \partial_\mu) \Omega^{-1}(x) 
 ,
 \label{V-ctransf}
\end{equation}
while $\mathscr{X}_\mu$ transforms like an adjoint matter field:\begin{equation}
  \mathscr{X}_\mu(x)   \rightarrow \mathscr{X}_\mu^\prime(x) = \Omega(x)  \mathscr{X}_\mu(x) \Omega^{-1}(x) 
 \label{X-ctransf}
 . 
\end{equation}

In the decomposition (\ref{decomp}), we introduce a new field 
\begin{equation}
 \bm{n}(x)=n^A(x)T_A ,  
\end{equation}
with a unit length in the sense that $n^A(x)n^A(x)=1$, which we call the \textit{color field}.  
In the decomposition (\ref{decomp}), the color field $\bm{n}(x)$ plays a crucial role as follows.
The color field is defined by the following property.
It must be a functional or composite operator of the original Yang-Mills field  $\mathscr{A}_\mu(x)$ such that it transforms according to the adjoint representation under the gauge transformation:  
\begin{equation}
  \bm{n}(x)   \rightarrow \bm{n}^\prime(x) = \Omega(x)  \bm{n}(x) \Omega^{-1}(x) 
   .
 \label{n-ctransf}
\end{equation}

The color field plays the key role in the reformulation.  Once a color field is given, the decomposition is uniquely determined by solving a set of defining equations and hence  $\mathscr{V}_\mu(x)$ and $\mathscr{X}_\mu(x)$ are written in terms of $\mathscr{A}_\mu(x)$ and $\bm{n}(x)$.
For $G=SU(2)$,  the defining equations are given by

(I) covariant constantness of color field $\bm{n}(x)$ in $\mathscr{V}_\mu(x)$:
\begin{equation}
  0 = D_\mu[\mathscr{V}]\bm{n}(x)   ,
\end{equation}

\text{(II) orthogonality of  $\mathscr{X}_\mu(x)$ to $\bm{n}(x)$:}
\begin{equation}
  0 = \mathscr{X}_\mu(x) \cdot \bm{n}(x)    .
\end{equation}
Then the decomposition for $G=SU(2)$ is uniquely determined as
\begin{align}
  & \mathscr{V}_\mu(x)= c_\mu(x)\bm{n}(x) +    ig^{-1} [ \bm{n}(x) , \partial_\mu \bm{n}(x) ]  ,
\nonumber\\
 & \quad\quad\quad\quad  c_\mu(x) :=  \mathscr{A}_\mu(x)  \cdot \bm{n}(x) ,
\nonumber\\
 & \mathscr{X}_\mu(x) =   i g^{-1} [ D_\mu[\mathscr{A}] \bm{n}(x) , \bm{n}(x) ] ,
\end{align}

To arrive at the result (\ref{QCDaction2}),  we have used the following facts. See Appendix~\ref{app:reformulation} for the details.

(i) 
The $O(\mathscr{X})$ terms vanish,  
$\frac{1}{2} \mathscr{F}^{\mu\nu}[\mathscr{V}] \cdot (D_\mu[\mathscr{V}] \mathscr{X}_\nu - D_\nu[\mathscr{V}] \mathscr{X}_\mu)=0$,
from the property of the new variables  as shown using the \textit{defining equations of the decomposition} (\ref{decomp}). 
This is somewhat similar to the usual background field method in which $O(\mathscr{X})$ terms in the quantum fluctuation field $\mathscr{X}_{\mu}$ are eliminated by requiring that the background field $\mathscr{V}_\mu$ satisfies the classical Yang-Mills equation of motion, i.e., $D_\mu[\mathscr{V}]\mathscr{F}^{\mu\nu}[\mathscr{V}]=0$. 
In our case,  however, $\mathscr{V}_\mu$ do not necessarily satisfy the classical equation of motion.

(ii)  To obtain $Q_{\mu\nu}^{AB}[\mathscr{V}]$ in (\ref{Q}), an $O(\mathscr{X}^2)$ term is eliminated, $-\frac{1}{2} \mathscr{X}^{\mu A}  D_\mu^{AC}[\mathscr{V}]D_\nu^{CB}[\mathscr{V}] \mathscr{X}^{\nu B}=0$,
by imposing  the condition:
\begin{equation}
 D_\mu[\mathscr{V}] \mathscr{X}^\mu=0 .
 \label{nMAG}
\end{equation}
For the reformulated QCD to be equivalent to the original QCD, we must impose such a constraint to avoid mismatch in the independent degrees of freedom, which is called the \textit{reduction condition} \cite{KMS06,KSM08}. 

(iii) The $O(\mathscr{X}^3)$ term vanishes, 
$
\frac{1}{2}  (D_\mu[\mathscr{V}] \mathscr{X}_\nu - D_\nu[\mathscr{V}] \mathscr{X}_\mu) \cdot ig[ \mathscr{X}^\mu , \mathscr{X}^\nu ] = 0 
$, since $D_\mu[\mathscr{V}] \mathscr{X}_\nu - D_\nu[\mathscr{V}] \mathscr{X}_\mu$ is orthogonal to $[ \mathscr{X}^\mu , \mathscr{X}^\nu ]$.

For $G=SU(2)$, $\mathscr{V}$ can be chosen in such a way that the field strength $\mathscr{F}[\mathscr{V}]$ of the field $\mathscr{V}$ is proportional to $\bm{n}$:
\begin{align}
 \mathscr{F}_{\mu\nu}[\mathscr{V}](x) 
:=& \partial_\mu \mathscr{V}_\nu(x)  - \partial_\nu \mathscr{V}_\mu(x)   - ig [\mathscr{V}_\mu(x) ,  \mathscr{V}_\nu(x)]
\nonumber\\
=&  \bm{n}(x) G_{\mu\nu}(x) ,
\label{F=nG}
\end{align}
where $G_{\mu\nu}$ is a gauge--invariant antisymmetric tensor of rank 2, i.e., 
$\mathscr{F}_{\mu\nu}^\prime[\mathscr{V}](x)=\mathscr{F}_{\mu\nu}[\mathscr{V}^\prime](x)=\Omega(x)  \mathscr{F}_{\mu\nu}[\mathscr{V}](x) \Omega^{-1}(x) = \bm{n}^\prime(x) G_{\mu\nu}(x) 
$.  The explicit form of $G_{\mu\nu}$ is written in term of $\mathscr{A}_\mu(x)$ and $\bm{n}(x)$ as
\begin{align}
 G_{\mu\nu}(x)  &=    \partial_\mu [ \bm{n}(x) \cdot \mathscr{A}_\nu(x)] - \partial_\nu [ \bm{n}(x) \cdot \mathscr{A}_\mu(x)] 
\nonumber\\&
+   i g^{-1} \bm{n}(x) \cdot [\partial_\mu \bm{n}(x) , \partial_\nu \bm{n}(x)  ] .
\end{align}

In the present approach, we wish to regard the field decomposition as a change of variable from the original gluon field to new variables describing a reformulated Yang-Mills theory in the quantum level \cite{KMS06,Kondo06,KSM08} (see \cite{KKMSS05,KKMSSI05,IKKMSS06,SKKMSI07,SKKMSI07b,KSSMKI08,SKKISF09,SKS09} for the corresponding lattice gauge formulation).  To achieve this goal, first of all, $\bm{n}(x)$ must be written as a functional of $\mathscr{A}_\mu(x)$ and thereby all new fields are written in terms of the original gluon field  $\mathscr{A}_\mu(x)$.   
Such a required relationship between $\mathscr{A}_\mu(x)$ and  $\bm{n}(x)$ is given by the reduction condition which is given as a variational problem of obtaining an absolute minimum of a given functional.  
The condition for local minima is given in the form of a differential equation. For $G=SU(2)$, 
\begin{equation}
 [ \bm{n}(x) , D_\mu[\mathscr{A}]D_\mu[\mathscr{A}]\bm{n}(x) ] =0 .
 \label{dRed}
\end{equation}
This is another form of (\ref{nMAG}). 
See \cite{KMS06} in $SU(2)$ case and \cite{KSM08} in $SU(N)$ case for the full details. 

Remarkable properties of new variables are as follows. 
First, we remind you of the role played by the field $\mathscr{V}$.

$\bullet$ The variable $\mathscr{V}_\mu$ alone is responsible for the Wilson loop  operator $W_C[\mathscr{A}]$ and the Polyakov loop operator $L[\mathscr{A}]$ in the sense that  
\begin{equation}
W_C[\mathscr{A}]=W_C[\mathscr{V}] , \quad
L[\mathscr{A}]=L[\mathscr{V}] .
\end{equation}
where the Wilson loop  operator is defined by
\begin{equation}
 W_C[A] := \mathcal{N}^{-1} {\rm tr} \left[ \mathscr{P} \exp \left\{ ig \oint_C dx^\mu \mathscr{A}_\mu(x) \right\} \right] ,
\end{equation}
where $\mathscr{P}$ denotes the path ordering and the normalization factor $\mathcal{N}$ is the dimension of the representation $R$, in which the Wilson loop is considered, i.e.,  
$
 \mathscr{N}:=d_R = {\rm dim}({\bf 1}_R) = {\rm tr}({\bf 1}_R)  
$.
The Polyakov loop operator will be defined later. 
In other words, $\mathscr{X}_\mu$ do not contribute to the Wilson loop and the Polyakov loop in the operator level. 
This is because the defining equation for the decomposition is a  (necessary and) sufficient condition for a \textit{gauge-invariant  Abelian dominance} (or $\mathscr{V}$ dominance) in the operator level. 
This proposition was first proved in \cite{Cho00}  for $SU(2)$ and   for $SU(N)$ in the continuum \cite{Kondo08} and for $SU(N)$ on a lattice \cite{KS08}.  On the lattice, the equality does not exactly hold due to non-zero lattice spacing $\epsilon$, but the deviation vanishes in the continuum limit of the lattice spacing $\epsilon$ going to zero, $\epsilon \rightarrow 0$.
It should be remarked that both the Wilson loop operator and the Polyakov loop operator are gauge-invariant quantities and that their average do not depend on the gauge fixing condition adopted in the calculation. 
 
 $\bullet$ We can introduce a gauge-invariant magnetic monopole current $k$ in Yang-Mills theory (without matter fields) where $k$ is the ($D-3$)-form.  For $D=4$ and $G=SU(2)$, 
\begin{align}
 k_\mu(x) :=& \partial_\nu {}^*G_{\mu\nu}(x) ,
\\
 G_{\mu\nu} :=&  \bm{n} \cdot \mathscr{F}_{\mu\nu}[\mathscr{V}]
  =  \partial_\mu c_\nu - \partial_\nu c_\mu +i g^{-1} \bm{n} \cdot [\partial_\mu \bm{n} ,  \partial_\nu \bm{n} ] ,
  \nonumber
\end{align}
where $f_{\mu\nu}$ is gauge-invariant field strength. 
This is because the field strength   $\mathscr{F}_{\mu\nu}[\mathscr{V}]:= \partial_\mu \mathscr{V}_\nu - \partial_\nu \mathscr{V}_\mu -ig [\mathscr{V}_\mu , \mathscr{V}_\nu ]$ is proportional to $\bm{n}$:
\begin{equation}
 \mathscr{F}_{\mu\nu}[\mathscr{V}]
 = \bm{n} \{ \partial_\mu c_\nu - \partial_\nu c_\mu +i g^{-1} \bm{n} \cdot [\partial_\mu \bm{n} ,  \partial_\nu \bm{n} ] \}  .
\end{equation}
 $\bullet$ The gauge-invariant ``Abelian" dominance (or $\mathscr{V}$ dominance) and magnetic monopole dominance (constructed from $\mathscr{V}$) in quark confinement have been confirmed at $T=0$ (and $\mu_q=0$)  by comparing string tensions calculated from the Wilson loop average by numerical simulations by \cite{IKKMSS06} for SU(2) and by \cite{SKKMSI07b} for SU(3). 
Here it should be remarked that the ``Abelian" dominance is the dominance for the vacuum expectation value (or average):
\begin{equation}
\langle W_C[\mathscr{A}] \rangle \simeq \langle W_C[\mathscr{V}] \rangle 
, \quad 
\langle  L[\mathscr{A}] \rangle  \simeq  \langle  L[\mathscr{V}] \rangle.
\end{equation}

Next, we pay attention to the role played by the remaining field $\mathscr{X}$. 
\\
$\bullet$ In the absence of dynamical quarks (corresponding to the limit $m_q=\infty$ of QCD, i.e., gluodynamics), 
$\mathscr{X}_\mu^A$ decouple in the low-energy regime as  the correlator $\left< \mathscr{X}_\mu^A(x) \mathscr{X}_\mu^A(y) \right>$ behaves like a massive correlator with mass $M_X$. 
In fact, numerical simulations demonstrate for $G=SU(2)$ and $D=4$ \cite{SKKMSI07} 
\begin{equation}
M_X=1.2 \sim 1.3 \ {\rm GeV}.
\label{M_X}
\end{equation}
We can understand this result as follows.
The field $\mathscr{X}_\mu$  can acquire the (gauge-invariant) mass dynamically. This comes from a fact that, in sharp contrast to the field $\mathscr{A}_\mu$, 
 a ``gauge-invariant mass term" for $\mathscr{X}_\mu$ can be introduced
\begin{equation}
\frac12 M_X^2 \mathscr{X}_\mu^A(x) \mathscr{X}_\mu^A(x) ,
\end{equation}
since $\mathscr{X}_\mu^A(x) \mathscr{X}_\mu^A(x)$ is a gauge-invariant operator.
Moreover, this mass term can originate from a vacuum condensation  of ``mass dimension-2", $\left< \mathscr{X}_\nu^B(x) \mathscr{X}_\nu^B(x) \right> \not= 0$ as proposed in \cite{Kondo01}.
In fact, this condensation can be generated through self-interactions $O(\mathscr{X}^4)$ among  $\mathscr{X}_\mu$ gluons, $M_X^2 \simeq \left< \mathscr{X}_\nu^B(x) \mathscr{X}_\nu^B(x) \right>$, as examined in \cite{Kondo06,KKMSS05}.
It is instructive to remark that the value (\ref{M_X}) agrees with the earlier result of the off-diagonal ``gluon mass'' $M_A$ in the Maximally Abelian (MA) gauge \cite{AS99} for $SU(2)$ case, $M_A \simeq 1.2$ GeV. See \cite{SAIIMT02} for $SU(3)$ case, $M_A \simeq 1.1$ GeV.
In MA gauge, it was shown that even at finite temperatures Abelian dominance (diagonal part dominance and off-diagonal part suppression) holds for the spatial propagation of gluons in the long distance greater than 0.4fm.  It was observed that the diagonal gluon correlator largely changes between the confinement and the deconfinement phase, while the off-diagonal gluon correlator is almost the same even in the deconfinement phase \cite{AS00}. 
Although the similar results are expected to hold in  our formulation, this observation must be checked directly, as will be confirmed in \cite{KSSK10}.

$\bullet$ In the presence of  dynamical quarks ($m_q<\infty$), $\mathscr{X}_\mu^A$ is responsible for chiral-symmetry breaking in the following sense.   
We consider to integrate out the field $\mathscr{X}_\mu^A$ in a naive way. 
This helps us to obtain an intuitive and qualitative understanding for the interplay between the chiral symmetry breaking and confinement. 
Later, this integration procedure will be reconsidered from the viewpoint of the renormalization group to obtain a systematic improvement of the result. 

Here we neglect $O(\mathscr{X}^3)$ and $O(\mathscr{X}^4)$ terms, which will be taken into account later. Then the integration over $\mathscr{X}_\mu^A$ can be achieved by the Gaussian integration according to \cite{Kondo06}. 
Consequently, a nonlocal 4 fermion-interaction is generated:
\begin{align}
 S_{\rm eff}^{\rm QCD} =& S_{\rm eff}^{\rm glue} + S_{\rm eff}^{\rm gNJL} ,
\nonumber\\
 S_{\rm eff}^{\rm glue} :=& \int d^Dx  \frac{-1}{4} (\mathscr{F}_{\mu\nu}[\mathscr{V}])^2   
\nonumber\\&
+ \frac{i}{2} \ln \det Q[\mathscr{V}]_{\mu\nu}^{AB} - i \ln \det G[\mathscr{V}]^{AB} 
 ,
\nonumber\\
S_{\rm eff}^{\rm gNJL} :=&  \int d^Dx \ \bar{\psi} (i\gamma^\mu \mathcal{D}_\mu[\mathscr{V}]  -\hat{m}_q + i  \gamma^0 \mu) \psi 
\nonumber\\ 
 +& \int d^Dx \int d^Dy  \frac{g^2}{2}  \mathscr{J}^{\mu}_{A}(x)  Q^{-1}[\mathscr{V}]_{\mu\nu}^{AB}(x,y) \mathscr{J}^{\nu}_{B}(y)  
 ,
\end{align}
where the last term $- \ln \det G^{AB}$ in $S_{\rm eff}^{\rm glue}$ comes from the Faddeev-Popov determinant associated with the reduction condition (\ref{nMAG}) (see \cite{KMS05} for the precise form). 

This is a nonlocal version of a gauged NJL model (realized after Fierz transformation).   
The chiral-symmetry breaking/restoration transition and the phase structure of a local version of  gauge NJL models were first studied by solving the Schwinger-Dyson equation in the ladder approximation for QED-like \cite{Miransky85,KMY89,KKM89} (see \cite{Kondo91} for a review) and QCD-like \cite{KSY91} running gauge coupling constant. 
They are confirmed later by a systematic approach of the   renormalization group \cite{Aoki-etal99}.

The range of the nonlocality is determined by the correlation length $\xi$, which is characteristic of the color exchange through gluon fields. 
Therefore, this correlation length $\xi$ is identified with the inverse of the effective mass $M_X$, i.e., $\xi \simeq M_X^{-1}$. In fact, $(Q^{-1})_{\mu\nu}^{AB}(x,y)$ is the $\mathscr{X}$ field correlator, see (\ref{QCDaction2}).

In other words, $M_X$ is identified with the ultraviolet cutoff $\Lambda$ below which the effective NJL model appears and works well. Interesting enough, $M_X$ is nearly equal to the ultraviolet cutoff adopted in the NJL model 
\begin{equation}
\sqrt{p^2} \lesssim \Lambda_4=1.4 {\rm GeV} , \quad |\bm{p}| \lesssim  \Lambda_3=0.6 {\rm GeV} , 
\end{equation}
see \cite{HK94}.

We can decompose the gauge field $\mathscr{A}_\mu$ into the the low-energy (light) mode $p<M_X$ and high-energy (heavy) mode $p>M_X$:  
\begin{align}
 \mathscr{A}_\mu(p)
=&\mathscr{A}_\mu(p) \theta(M_X^2-p^2)+\mathscr{A}_\mu(x) \theta(p^2-M_X^2)
  .
  \label{decomp2}
\end{align}
In the above treatment, $\mathscr{X}_\mu(p)$ is supposed to have only the high-energy mode. 
The low-energy mode, if any, will be responsible for the vacuum condensation \cite{Kondo06}. 
For the precise understanding, we need the renormalization group treatment as given later and the implications for the nonlocal NJL model will be discussed there.

\section{Gluon sector and gauge fixing}\label{sec:gauge}

The Polyakov loop operator $L$ and the chiral operator $\bar\psi \psi$ are gauge-invariant quantities.  Therefore, their average do not depend on the gauge-fixing procedure adopted in the calculation. 
We can choose a gauge in which the actual calculation becomes easier than other gauges. 

In what follows, we  treat the time-component $\mathscr{V}_0$ and space-component $\mathscr{V}_j$ of $\mathscr{V}_\mu$ differently to consider the finite-temperature case.  
We consider the following Polyakov gauge modified for new variables in our reformulation. 
If the color field $n^A(x)$ is uniform in time, 
\begin{equation}
  \partial_0 n^A(x) = 0  \Leftrightarrow
n^A(x) = n^A (\bm{x})
 , 
 \label{my-gauge-0}
\end{equation}
then $\mathscr{V}_0$ reduces to
\begin{equation}
 \mathscr{V}_0^A(x) = c_0(x)n^A (\bm{x})  \quad (A=1,2,3)
 .
 \label{my-gauge-1}
\end{equation}
Moreover, if  $c_0(x)$ is uniform in time, 
\begin{equation}
  \partial_0 c_0(x) = 0  \Leftrightarrow
c_0(x) = c_0(\bm{x})
 , 
\end{equation}
then $\mathscr{V}_0$ reduces to
\begin{equation}
 \mathscr{V}_0^A(x) = c_0(\bm{x})n^A (\bm{x})  \quad (A=1,2,3)
 ,
 \label{my-gauge-2}
\end{equation}
which satisfies 
\begin{equation}
 \partial_0 \mathscr{V}_0^A(x) = 0  \quad (A=1,2,3)
 .
 \label{my-gauge-3}
\end{equation}
In this setting, $\mathscr{V}_j$ are given by
\begin{equation}
 \mathscr{V}_j^A(x) = c_j(x)n^A(\bm{x}) + g^{-1} \epsilon^{ABC} \partial_j n^B(\bm{x}) n^C(\bm{x}) 
 . 
\end{equation}

(1) In order to simplify the calculation of the Polyakov line, we adopt the Polyakov gauge in which the gauge field is diagonal and time-independent: 
for the background field $\mathscr{V}_0^A(x)$, 
\begin{equation}
 \mathscr{V}_0^A(x) = c_0(\bm{x}) \delta^{A3}  ,
\end{equation}
which leads to
\begin{equation}
 \partial_0 \mathscr{V}_0^A(x) = 0 
 .
\end{equation}
This is realized, if we take the gauge
\footnote{This is an oversimplified choice for the color field $\bm{n}(x)$.  
By this choice, we can not separate the non-perturbative contribution coming from topological configurations such as magnetic monopole.
It is desirable to take into account color field degrees of  freedom explicitly to see the effect of magnetic monopole in the confinement/deconfinement transition. 
}
\begin{equation}
  n^A(\bm{x}) = \delta^{A3} 
 . 
\end{equation}
In this gauge, the space-component reads
\begin{equation}
 \mathscr{V}_j^A(x) = c_j(x)\delta^{A3}  
 . 
\end{equation}
which is not time-independent, $\partial_0 \mathscr{V}_j^A(x) \not= 0$.

(2) We expand the theory around the non-trivial uniform background $g^{-1}T\varphi  \delta^{A3} $ for the time-component $\mathscr{V}_0^A$, while the trivial background for  space-components  $\mathscr{V}_j^A$:\footnote{
I have assumed that the spatial component of $\mathscr{V}_\mu$ has a trivial background. 
In view of logical consistency, one must expand the spatial and temporal components around non-trivial backgrounds, and then one must search for the minima of the effective potential calculated as a function of two variables, i.e., the temporal and spatial backgrounds.  In this paper, it is assumed that a minimum is realized at vanishing spatial background and that the neglection of the spatial background does not so much affect the confinement/deconfinement transition temperature. 
Indeed, it must be checked whether this assumption is good or not.
}
\begin{align}
 \mathscr{V}_0^A(x) =& c_0(\bm{x})\delta^{A3} , \ c_0(\bm{x}) = g^{-1}T\varphi  + v_0(\bm{x}) 
 ,
\nonumber\\
 \mathscr{V}_j^A(x) =&  0+ v_j^A(x)
 , 
\end{align}
such that 
$\langle c_0(\bm{x}) \rangle = g^{-1}T \langle \varphi \rangle  + \langle v_0(\bm{x}) \rangle=g^{-1}T \langle   \varphi \rangle $ with $\langle v_0(\bm{x}) \rangle=0$ and $\langle v_j^A(x) \rangle=0$. 
Here the prefactor $g^{-1}T=(g\beta)^{-1}$ was introduced just for the purpose of simplifying the expression of the Polyakov loop, see (\ref{P}). 

(3) We take into account the expansion up to quadratic  in the fluctuation fields $v_0$ and $v_j$, which we call the quadratic approximation.

In the calculation of $Q_{\mu\nu}^{AB}$, if we neglect all fluctuation fields $v_0$ and $v_j$, namely,   
$\mathscr{V}_0^A (x) =  g^{-1}T \varphi \delta^{A3}$ and  $\mathscr{V}_j^A (x) =  0$, then we can put 
$\mathscr{F}_{\mu\nu}^{C}[\mathscr{V}] = 0$ in $Q_{\mu\nu}^{AB}$, and  $Q_{\mu\nu}^{AB}$ is diagonal in the Lorentz indices:
\begin{align} 
Q_{\mu\nu}^{AB}   &=   G^{AB} g_{\mu\nu} , 
  \label{approx0}
\\ 
G^{AB}  
& =  - \delta^{AB} \partial_\mu^2  +   (\delta^{AB}  - \delta^{A3}\delta^{B3})(T \varphi)^2 + 2  \epsilon^{AB3} T \varphi \partial_0  
 .
\nonumber
\end{align}
In this approximation, we have
\begin{align}
 & G^{AB} =   - \delta^{AB} \partial_\ell^2  - D_0^{AC}[\mathscr{V}]D_0^{CB}[\mathscr{V}] ,
  \label{approx}
\end{align}
with
\begin{align}
    - D_0^{AC}[\mathscr{V}]D_0^{CB}[\mathscr{V}]   
=&  -   \delta^{AB} \partial_0^2
+ 2  \epsilon^{AB3} T \varphi \partial_0  
 \nonumber\\&
+ (\delta^{AB}  - \delta^{A3}\delta^{B3})(T \varphi)^2 
  .
\end{align}
Thus we rewrite the gluon part $S_{\rm eff}^{\rm glue}[\mathscr{V}]$ as
\begin{align}
 & S_{\rm eff}^{\rm glue}[\mathscr{V}]  
 \nonumber\\ 
&=    \frac12 \beta \int d^3x \mathscr{V}_0(\bm{x})  (-\partial_j \partial_j)  \mathscr{V}_0(\bm{x})   
 \nonumber\\&
 + \frac12  \int d^4x  \mathscr{V}_{T}^A(x)  \{  
    -  \delta^{AB}  \partial_\ell^2   - \delta^{AB}  \partial_0^2
   \} \mathscr{V}_{T}^B(x) 
 \nonumber\\&
 + \frac12 \int d^4x \mathscr{V}_{L}^A(x)  \{    - \delta^{AB}  \partial_0^2  \} \mathscr{V}_{L}^B(x) 
  ,
\nonumber\\&
+ \frac{i}{2} \ln \det Q[\mathscr{V}]_{\mu\nu}^{AB} - i \ln \det G[\mathscr{V}]^{AB} 
 ,
  \label{mode-decomp1}
\end{align}
where $\beta$ is the inverse temperature $\beta:=1/T$, and $\mathscr{V}_{T}$ and $\mathscr{V}_{L}$ denote the transverse and longitudinal components of $\mathscr{V}_\mu$ respectively.

\section{Polyakov loop}\label{sec:Polyakov}

For $G=SU(2)$, the Polyakov loop operator $L(\bm{x})=L[\mathscr{V}_0(\bm{x}, \cdot)]$ is defined by
\begin{align}
L(\bm{x}) :=&   \frac12 {\rm tr}(P) ,
\nonumber\\
 P(\bm{x}) :=& \mathscr{P} \exp \left[ ig \int_{0}^{\beta=1/T} dx_0 \mathscr{V}_0^A(\bm{x},x_0)  \frac{\sigma_A}{2} \right]  
, 
\label{P}
\end{align}
where 
$P P^\dagger = \mathbf{1}$
and
$\det P =1$.
In the above gauge choice, 
\begin{equation}
 P(\bm{x}) =   \exp \left[ ig \beta c_0(\bm{x})  n^A(\bm{x}) \frac{\sigma_A}{2}   \right] . 
 \nonumber
\end{equation}
After a suitable ($t$-independent) gauge transformation, the color field $n^A(\bm{x})$ is eliminated:
\begin{equation}
 L(\bm{x}) 
=  \frac12 {\rm tr}(\exp \left[ ig \beta c_0(\bm{x})   \frac{\sigma_3}{2}   \right] )
= \cos \left( \frac{g \beta c_0(\bm{x})}{2}    \right) .  
\label{LL}
\end{equation}

Owing to periodicity and center symmetry, we can restrict the Polyakov loop average to $\langle L \rangle \ge 0$ for $G=SU(2)$.  Then the Polyakov loop average $\langle L[\mathscr{V}] \rangle$ is bounded from above by $L[ \langle \mathscr{V}_0(\bm{x}, \cdot) \rangle]$:
\begin{equation}
 0 \le \langle L[\mathscr{V}_0(\bm{x}, \cdot)] \rangle
\le     L[ \langle \mathscr{V}_0(\bm{x}, \cdot) \rangle]  
= \cos \left(  \frac{\langle \varphi \rangle}{2}    \right)  ,
\end{equation}
where we have only to consider the range $0 \le \varphi \le \pi$.
This inequality follows from the Jensen inequality, since $\cos (x)$ is concave for $0 \le x\le \pi/2$, see \cite{MP08}.

In the case of $m_q=\infty$, if the center-symmetry is broken $\langle L \rangle >0$, namely, deconfinement  takes place, then  the vacuum (as a minimum of the effective potential $V_{\rm eff}(\varphi)$) is realized at $\langle \varphi \rangle < \pi$.
If the vacuum is realized at $\langle \varphi \rangle=\pi$, then the center-symmetry is restored $\langle L \rangle =0$, namely,  confinement  occurs.   The relation (\ref{LL}) yields the relationship for the average between the gauge field and the Polyakov loop operator:  
\begin{equation}
  \langle  \arccos L(\bm{x}) \rangle = \frac{g \beta \langle c_0(\bm{x}) \rangle  }{2}  = \frac{\langle \varphi \rangle}{2} ,
\end{equation}
where the left-hand side is the average of an gauge-invariant object (since $L$ is gauge invariant) and happens to agree with the average $\langle  \mathscr{V}_0^3 \rangle$ of the gauge field in the Polyakov gauge. 
It is also shown \cite{MP08} that the converse is  true: In the center-symmetry-restored phase, $\langle \varphi \rangle=\pi$, since 
\begin{equation}
   \frac{\langle \varphi \rangle}{2}  =  \langle  \arccos L(\bm{x}) \rangle =  \arccos \langle L(\bm{x}) \rangle   = \frac{\pi}{2} .
\end{equation}

Therefore, $\langle \mathscr{V}_0 \rangle$ or $\langle \varphi \rangle$ in the Polyakov gauge gives a direct physical interpretation as an order parameter for the confinement/deconfinement (order-disorder) phase transition. 
The effective potential $U_{\rm eff}(\langle L \rangle)$ of the Polyakov loop average $\langle L \rangle$  could be different from the effective potential $V_{\rm eff}(\langle  \mathscr{V}_0 \rangle)$ of the gauge field average $\langle  \mathscr{V}_0 \rangle$ in the following sense.
Although both potentials give the same critical temperature $T_d$ as a boundary between $\langle L  \rangle=0$ and $\langle L \rangle \not= 0$,  
the value of the effective potential $U_{\rm eff}(\langle L[\mathscr{V}_0 ] \rangle)$ does not necessarily agree with $U_{\rm eff}(L[ \langle \mathscr{V}_0  \rangle])=V_{\rm eff}(\langle  \mathscr{V}_0 \rangle)$ at a given temperature $T$, since we have only an inequality 
$
\langle L[\mathscr{V}_0 ] \rangle \le L[ \langle \mathscr{V}_0  \rangle]
$.
This difference could affect the critical exponent and other physical quantities of interest. Therefore, the result obtained from $V_{\rm eff}(\langle  \mathscr{V}_0 \rangle)$ must be carefully examined.

\section{Deriving the confinement/deconfinement transition}\label{sec:deconfinement}

In this section, we restrict our consideration to the pure glue case. 
We show that the pure gluon part $S_{eff}^{\rm glue}$ can  describe confinement/deconfinement transition signaled by the Polyakov loop average $\langle L \rangle$. 
In this section, we completely follow two remarkable papers by Marhauser and Pawlowski \cite{MP08} and by Braun, Gies and  Pawlowski \cite{BGP07}, which succeeded to show the transition for the first time based on the functional renormalization group (FRG).
In the next section, we explain how these results are understood from the Landau-Ginzburg argument. 

We consider the flow equation called the Wetterich equation \cite{Wetterich93} for the $k$(RG scale)-dependent effective action $\Gamma_k$:
\begin{align}
\partial_t \Gamma_k[\Phi] 
=& \frac12 {\rm STr} \left\{ \left[ \frac{\overrightarrow{\delta}}{\delta \Phi^\dagger} \Gamma_k[\Phi] \frac{\overleftarrow{\delta}}{\delta \Phi} + R_{\Phi,k} \right]^{-1} \cdot \partial_t R_{\Phi,k} \right\}
 ,
\end{align}
where $t$ is the RG time 
$
  t := \ln \frac{k}{\Lambda}
$, 
$
 \partial_t := \frac{\partial}{\partial t} = k \frac{d}{dk} 
$
for some reference scale (UV cutoff) $\Lambda$ 
and $R_{\Phi,k}$ is the regulator function for the field $\Phi$. 
Here ${\rm STr}$ denotes the super-trace introduced to include both  commutative field (gluon) and anticommutative field (quark, ghost). 
See \cite{BTW00,Gies06} for  reviews of the functional renormalization group. 

If we restrict our consideration to the pure glue case $S_{\rm YM}$ under the gauge $n^A(x)=\delta^{A3}$, then the relevant fields $\Phi$ are $\mathscr{V}_\mu^A(x)$ , $\mathscr{X}_\mu^A(x)$ and FP ghosts (ghost and antighost) $\mathscr{C}^A(x), \bar{\mathscr{C}}^A(x)$, i.e., $\Phi^\dagger = (\mathscr{V}_\mu^A,\mathscr{X}_\mu^A,\mathscr{C}^A,\bar{\mathscr{C}}^A)$. 
In this section, we use the Euclidean formulation. 
In the modified Polyakov gauge and within the quadratic approximation adopted in sec.~\ref{sec:gauge}, 
\begin{align}
 \partial_t \Gamma_k
 =& \frac12 {\rm Tr} \left\{ \left[ \frac{\overrightarrow{\delta}}{\delta \mathscr{V}^\dagger} \Gamma_k \frac{\overleftarrow{\delta}}{\delta \mathscr{V}} + R_{k} \right]^{-1} \cdot \partial_t R_{k} \right\}
\nonumber\\&
 + \partial_t \frac12 {\rm Tr}  \{ \ln [Q_{\mu\nu}^{AB}+\delta^{AB}\delta_{\mu\nu}R_{k} ] \} 
\nonumber\\&
 - \partial_t   {\rm Tr}  \{ \ln [G^{AB}+\delta^{AB}R_{k} ] \} 
  .
\end{align}
where the second contribution in the right-hand side comes from the $\mathscr{X}$ field and the last one from the ghosts fields \cite{KMS05}, and we have used the same regulator function $R_k$ for the gluon and ghost up to the difference due to the tensor structure.

We neglect back-reactions of the $\mathscr{V}_{0}$ potential on the other gauge fields $\mathscr{V}_{j}$, as in the treatment \cite{MP08}.  
Assuming an expansion around $\mathscr{V}_{j}=0$, $\Gamma_{k}^{(2)}:=\frac{\overrightarrow{\delta}}{\delta \mathscr{V}^\dagger} \Gamma_k \frac{\overleftarrow{\delta}}{\delta \mathscr{V}}$ is block-diagonal like the regulators, and  the flow equation can be decomposed into a sum of two contributions:
under the approximation (\ref{approx0}),
\begin{align}
 \partial_t \Gamma_k
 =& \frac12 {\rm Tr} \left[ \left( \frac{1}{\Gamma_k^{(2)}+R_{k}} \right)_{\mu\nu} \cdot \partial_t R_{k,\mu\nu} \right]
\nonumber\\&
 + \partial_t {\rm Tr}  \{ \ln [G^{AB}+\delta^{AB}R_{k} ] \}
  ,
\end{align}
where the gluon regulator $R_{k,\mu\nu}$ is a block-diagonal matrix in field space,
\begin{align}
  R_{k,00} =& R_{0,k} = Z_0 R_{\rm opt,k}(\bm{p}^2), \
  R_{k,0j} = 0 = R_{k,j0} ,
\nonumber\\
  R_{k,j\ell} =& R_{T,k} T_{j\ell}(\bm{p})
= Z_j T_{j\ell}(\bm{p}) R_{\rm opt, k_T}(\bm{p}^2) ,
\end{align}
where $T_{j\ell} := \delta_{j\ell} - \frac{p_jp_\ell}{p_m^2}$ is the transverse projection operator and 
$R_{\rm opt,k}(\bm{p}^2)$ is the (3 dim.) optimized choice \cite{Litim00}:
\begin{equation}
 R_{\rm opt,k}(\bm{p}^2) = (k^2-\bm{p}^2) \theta(k^2-\bm{p}^2) .
\end{equation}

The first term in the right-hand side encodes the quantum fluctuations of $\mathscr{V}_0$, while the second one encodes those of the other components of the gauge field and ghosts.  
In the present truncation, the second term is a total derivative with respect to $t$, and does not receive contributions from the first term.  Therefore, we can evaluate the flow of the second contribution, and use its output $V_{T,k}(\mathscr{V}_0)$ as an input for the remaining flow. 
\begin{align}
 \partial_t \Gamma_k
 =& \frac12 \beta \int \frac{d^3p}{(2\pi)^3} \left[ \left( \frac{1}{\Gamma_k^{(2)}
+R_\mathscr{V}} \right)_{00} \partial_t R_{0,k} \right]
\nonumber\\&
 +   \partial_t V_{T,k} 
  ,
\end{align}
where for $\omega=2\pi Tn$
\begin{align}
& V_{T,k} 
 :=    {\rm Tr}  \{  \ln [G^{AB}+\delta^{AB}R_{k} ] \} 
 \nonumber\\
=&    T \sum_{n \in \mathbb{Z}}  \int \frac{d^3p}{(2\pi)^3}  {\rm tr}\ln [\tilde{G}^{AB}(\omega, \bm{p})
+\delta^{AB}(k_T^2-\bm{p}^2) \theta (k_T^2-\bm{p}^2)]
  \nonumber\\ 
 =&   T \sum_{n \in \mathbb{Z}}  \int \frac{d^3p}{(2\pi)^3}   {\rm tr}\ln [ \delta^{AB}\bm{p}^2-D_0^2 
+ \delta^{AB}(k_T^2-\bm{p}^2) \theta (k_T^2-\bm{p}^2) ]  
 \nonumber\\
 =&    T \sum_{n \in \mathbb{Z}}  4\pi \int_{0}^{k_T} \frac{dp p^2}{(2\pi)^3}   {\rm tr}\ln [ \delta^{AB}k_T^2 -D_0^2   ] 
 \nonumber\\&
-   T \sum_{n \in \mathbb{Z}}  4\pi \int_{0}^{k_T} \frac{dp p^2}{(2\pi)^3}   {\rm tr}\ln [ \delta^{AB}\bm{p}^2-D_0^2 ] 
 + V_W
  .
  \label{V_{T,k}}
\end{align}
Here we have introduced the Weiss potential $V_W$ which was obtained by one-loop calculation \cite{Weiss81}:
\begin{align}
V_W  =&   
      {\rm Tr}  \ln [G^{AB}] 
  \nonumber\\ 
 =&    T \sum_{n \in \mathbb{Z}}  \int \frac{d^3p}{(2\pi)^3}  {\rm tr} \ln [\tilde{G}^{AB}(p_0=\omega, \bm{p})]
  \nonumber\\ 
=&   
    T \sum_{n \in \mathbb{Z}}  \int \frac{d^3p}{(2\pi)^3}   {\rm tr}\ln [\bm{p}^2+(\omega + T\varphi)^2] 
  \nonumber\\&
 +   T \sum_{n \in \mathbb{Z}}  \int \frac{d^3p}{(2\pi)^3}  {\rm tr}\ln [\bm{p}^2+(\omega - T\varphi)^2]
  ,
\end{align}
where we have neglected the $\varphi$-independent (or $\mathscr{V}_0$ independent) contributions.

\begin{figure}[ptb]
\begin{center}
  \includegraphics[width=5.5cm]{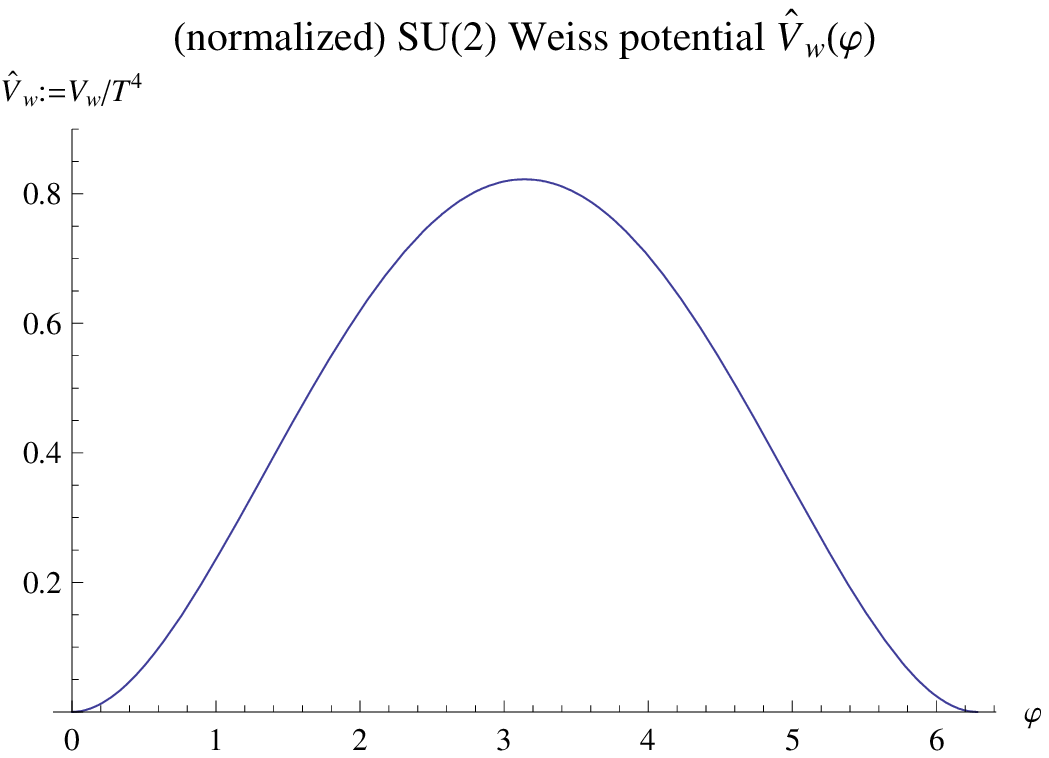}
  \includegraphics[width=5.5cm]{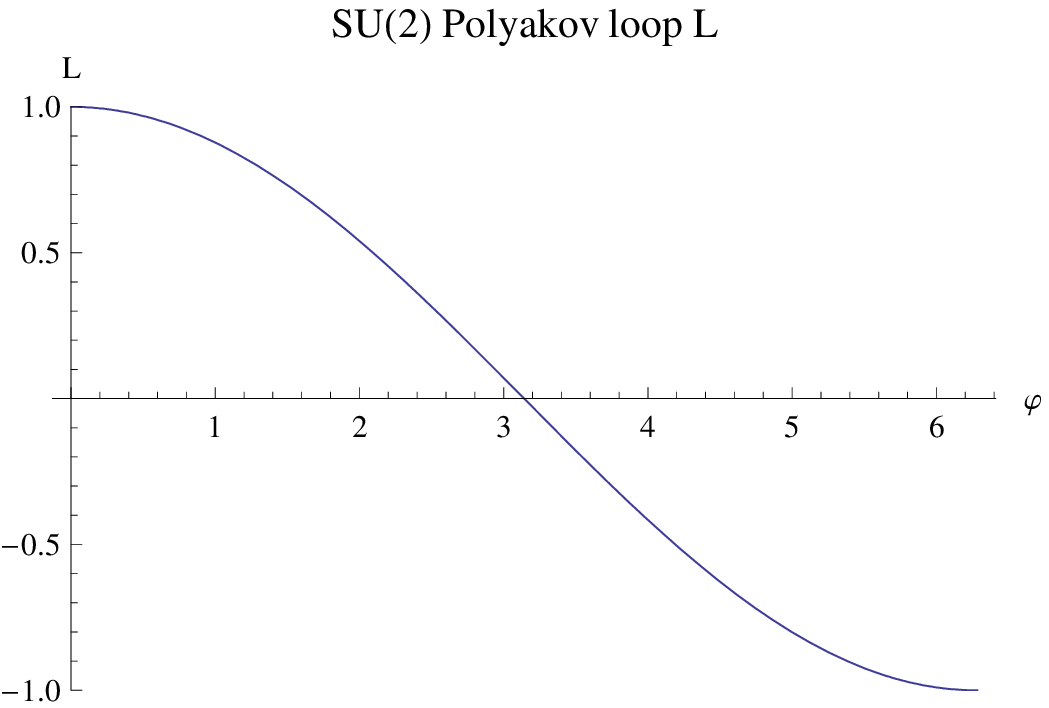}
\end{center}
  \caption{\small 
(The upper panel)  (normalized) SU(2) Weiss potential $\hat V_{W}$ as a function of $\varphi$.
(The lower panel) SU(2) Polyakov loop $L$ as a function of $\varphi$,
$L = \cos \left(  \frac{\varphi}{2}    \right) 
$.
}
  \label{fig:Weiss}
\end{figure}

\begin{figure}[ptb]
\begin{center}
  \includegraphics[width=6cm]{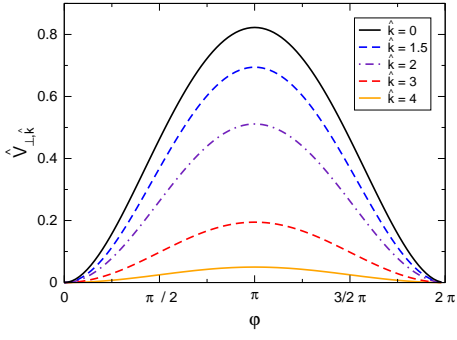}
\end{center}  
  \caption{$\hat V_{T,k}$ for different values of $\hat k$ [reprinted from \cite{MP08}].}
  \label{fig:VPreWeiss3D}
\end{figure}

The closed form of the Weiss potential is obtained after summing up the Matsubara frequencies:
\begin{align}
 V_W(\varphi)
=&    T^4 \left[ - \frac{1}{6}  (\varphi-\pi)^2 +  \frac{1}{12\pi^2}   (\varphi-\pi)^4   +   \frac{\pi^2}{12} \right]
\nonumber\\
 ({\rm mod} \ 2\pi)
 .
\end{align}
The Weiss potential $V_W$ is $g^2$ independent and the overall curve scales as $T^4$.
  $V_W$ has symmetries: 
$V_W(-\varphi)=V_W(\varphi)$ and $V_W(\varphi+2\pi n)=V_W(\varphi)$.
  $V_W(\varphi)$   has minima at $\varphi=2\pi n$, and the Polyakov loop has the nonvanishing value $L=\cos  \frac{\varphi}{2}=(- 1)^n$, implying deconfinement. See Fig.~\ref{fig:Weiss}.
  Therefore, $V_W(\varphi)$ is considered to be valid at very high temperature where the perturbation theory is trustworthy. 
  In  Fig.~\ref{fig:VPreWeiss3D}, we observe
\begin{equation}
\lim_{k \downarrow 0} V_{T,k} =  V_{T,0}  =  V_{W}   
  ,
\quad
\lim_{k \uparrow \infty} V_{T,k} =   0 .
\end{equation}

After integrating over the fields other than $\mathscr{V}_0$, we are lead to  the effective action of $\mathscr{V}_0$, 
\begin{align}
 \Gamma_{k}[\mathscr{V}_0] 
=&   \beta \int d^3x \left\{ -  \frac{1}{2}Z_0 \mathscr{V}_0(\bm{x})  \partial_j \partial_j  \mathscr{V}_0(\bm{x})    + V_{{\rm eff},k}^{\rm glue}[\mathscr{V}_0] \right\}  ,
 \nonumber\\ 
  V_{{\rm eff},k}^{\rm glue}[\mathscr{V}_0] =& V_{T,k}[\mathscr{V}_0] + \Delta V_k[\mathscr{V}_0]   .
  \label{V_g}
\end{align}
Then the flow equation is reformulated for $\Delta V_k$ with the external input $V_{T,k}$:
\begin{equation}
 \beta    \partial_{t}  (\Delta V_k[\mathscr{V}_0])
 = \frac12 \beta \int \frac{d^3p}{(2\pi)^3} \left[ \left( \frac{1}{\Gamma_k^{(2)}+R_{k}} \right)_{00} \partial_t R_{0,k} \right]
 ,
\end{equation}
where
\begin{equation}
\Gamma_{k}^{(2)}[\mathscr{V}_0] 
=   \beta  \left\{  Z_0   \bm{p}^2       + \partial_{\mathscr{V}_0}^2 V_{k}[\mathscr{V}_0] \right\}  
  .
\end{equation}

Using the specific regulator, 
$
  R_{0,k}
= Z_0 (k^2-\bm{p}^2)\theta(k^2-\bm{p}^2) 
$, 
which yields
\begin{align}
  \partial_t R_{0,k}
=  \left[ \partial_{t} Z_0   (k^2-\bm{p}^2)  + 2 Z_0  k^2 \right] \theta(k^2-\bm{p}^2)  
 ,
\end{align}
we can perform the momentum integration analytically. 
\begin{align}
 & \beta    \partial_{t}  (\Delta V_k[\mathscr{V}_0])    
\nonumber\\
=&  \frac23  \frac{1}{(2\pi)^2} \frac{  (\eta_k/5+1)  k^5 }{     Z_k^{-1} g^2 \beta^2 \partial_{\varphi}^2 (V_{T,k}[\mathscr{V}_0] + \Delta V_k[\mathscr{V}_0] )  + k^2   }
 ,
\end{align}
where we have introduced 
 the running coupling $\alpha_k$ defined by 
\begin{equation} 
 g_k^2 := Z_k^{-1} g^2, \quad \alpha_k := \frac{g_k^2}{4\pi} =  Z_k^{-1} \frac{g^2}{4\pi}
  ,
\end{equation}
and the anomalous dimension $\eta_k$  defined by
\begin{equation}
 \eta_k := \partial_{t} \ln Z_k = - \partial_{t} \ln \alpha_k .
\end{equation}

\begin{figure}[ptb]
\begin{center}
 \includegraphics[width=7.5cm]{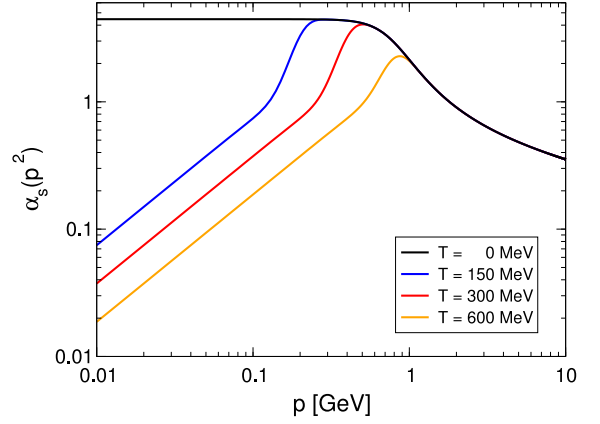}
\end{center} 
  \caption{The running gauge coupling constant $\alpha_s$ for temperatures $T=0,150,300,600$ MeV [reprinted from \cite{MP08}].}
  \label{fig:alpha}
\end{figure}

\begin{figure}[ptb]
\begin{center}
  \includegraphics[width=7.5cm]{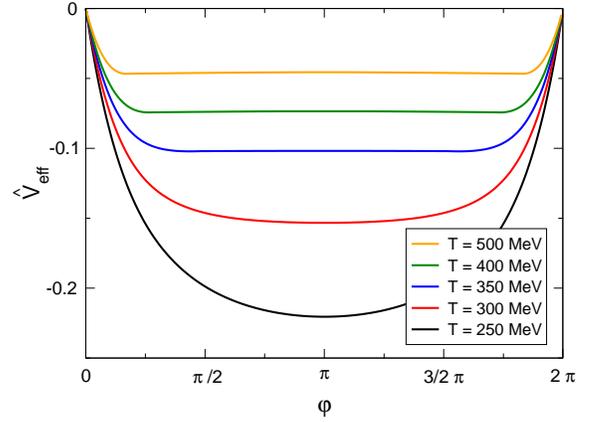}
\end{center}  
  \caption{Full effective potential $\hat V_{{\rm eff}}^{\rm glue}$, 
  normalized to 0 at $\varphi =0$ [reprinted from \cite{MP08}].}
  \label{fig:Veff}
\end{figure}

By introducing the dimensionless RG scale $\hat{k}$ and the dimensionless effective potential $\hat{V}$ defined by
\begin{equation}
 \hat{k} := \beta k = k/T  ,
 \quad 
 \hat{V} := \beta^4 V =V/T^4  
  ,
  \label{rescaling}
\end{equation}
the flow equation  is simplified as
\begin{equation}
     \partial_{\hat k}  \Delta \hat V_{\hat k}[\mathscr{V}_0]    
=    \frac{1}{6\pi^2} \frac{  (1+\eta_k/5 )  \hat k^2 }{   1+\frac{4\pi\alpha_k}{\hat k^2} \partial_{\varphi}^2 (\hat V_{T,\hat k}[\mathscr{V}_0] + \Delta \hat V_{\hat k}[\mathscr{V}_0] )     }
 ,
 \label{flow-eq}
\end{equation}
where all scales are measured in units of temperature. 
It turns out that the input in solving the flow equation is just a running gauge coupling constant $\alpha_k$.  A specific choice for the running gauge coupling constant is given in Fig.~\ref{fig:alpha}. For the derivation from the renormalization group, see \cite{BG05}. 
The flow is initialized in the broken phase at any temperature. 
By solving the flow equation in a numerical way with an input for the running gauge coupling given in Fig.~\ref{fig:alpha}, the full effective potential $\hat V_{\rm eff}$ (normalized to 0 at $\varphi =0$) is obtained in Fig.~\ref{fig:Veff} for various temperature.

According to \cite{MP08}, a second order phase transition occurs at a critical temperature
\begin{equation}
     T_d = 305_{-55}^{+40} \text{MeV}, \quad
T_d/\sqrt{\sigma}=0.69_{-.12}^{+.04}
 ,
\end{equation}
with the string tension $\sqrt{\sigma}=440$ MeV.
This agrees within errors with the lattice result
$T_d/\sqrt{\sigma}=0.709$.
Moreover, these results were confirmed by considering another gauge \cite{BGP07}.

\section{Understanding the existence of confinement transition according to the Landau-Ginzburg argument}\label{sec:Landau}

In this section, we show that some qualitative aspects of the deconfinement/confinement transition found in the previous section can be understood without detailed numerical works, although the precise value of the transition temperature $T_d$ cannot be determined without them. 

For $G=SU(2)$ in the pure Yang-Mills limit $m_q \rightarrow \infty$, the effective potential $V_{\rm glue}(L)$ for the Polyakov loop $L$  must be invariant under the center symmetry $Z(2)$.  Therefore, $V_{\rm glue}(L)$ is an even function of $L$, i.e., $V_{\rm glue}(L)=V_{\rm glue}(-L)$ where $L$ is real-valued $L=L^*$.  
Thus the Landau-Ginzburg argument suggests that the effective potential $V_{\rm eff}^{\rm glue}(L)$ for $G=SU(2)$ has the power-series expansion in  $L$  near the transition point $L=0$:  
\begin{equation}
 V_{\rm eff}^{\rm glue}(L) =  c_0 + \frac{c_2}{2} L^2   + \frac{c_4}{4}  L^4  + O(L^6)
 .
\end{equation}  
As the vacuum is specified as minima of the effective potential,  the confinement/deconfinement transition temperature $T_d$ is determined from the condition $c_2(T_d)=0$ so that the low-temperature ($T<T_d$) confinement phase $\langle L \rangle =0$ is realized for $c_2(T)>0$, while the high-temperature ($T>T_d$) deconfinement phase $\langle L \rangle \not=0$ is realized for $c_2(T)<0$, provided that the positivity $c_4(T)>0$ is maintained across the transition temperature. 
Consequently, the transition is of the 2nd order. 

Indeed, we confirm that the Landau-Ginzburg description is correct and valid for the confinement/deconfinement transition, by making use of the flow equation given in the previous section. 
This is a microscopic justification of the Landau-Ginzburg argument for the confinement/deconfinement transition.
In our treatment, however, it is more convenient to write the effective potential $V_{\rm eff}^{\rm glue}$ in terms of the angle variable $\varphi$ (rather than $L$) around the transition point $\varphi=\pi$ (instead of  $L=0$).
Defining $\tilde\varphi :=\varphi-\pi$, we find that 
$V_{\rm eff}^{\rm glue}(\tilde\varphi)$ must be an even function $V_{\rm eff}^{\rm glue}(\tilde\varphi)=V_{\rm eff}^{\rm glue}(-\tilde\varphi)$  due to the center symmetry  and hence odd terms (e.g., $\tilde\varphi$,  $\tilde\varphi^3$) do not appear: 
\begin{equation}
 V_{\rm eff}^{\rm glue}(\tilde\varphi)  =  C_{0} +  \frac{C_{2}}{2} \tilde \varphi^2 + \frac{C_{4}}{4!} \tilde \varphi^4 + O(\tilde \varphi^6)
  .
\end{equation}
At sufficiently high temperature, we observe that $C_2(T)<0$ and hence $V_{\rm eff}^{\rm glue}$ has the minimum at   $\tilde \varphi \not=0$ ($\Longleftrightarrow L\not=0$) leading to  deconfinement. 
In order to show the existence of the confinement/deconfinement transition at $T=T_d$,  $C_2(T)$ must change  the signature $C_2(T)>0$ below this  temperature $T<T_d$ and hence the minimum occurs at  $\tilde \varphi=0$ ($\Longleftrightarrow L=0$) leading to confinement.
Therefore, the confinement/deconfinement temperature $T_d$ is  determined by $C_2(T_d)=0$, provided that the positivity $C_4(T)>0$ is maintained.

For this purpose, we study the scale dependent effective potential $V_{\rm eff,k}^{\rm glue}$ at $k>0$
\begin{equation}
 V_{\rm eff,k}^{\rm glue}  =  C_{0,k} +  \frac{C_{2,k}}{2} \tilde \varphi^2 + \frac{C_{4,k}}{4!} \tilde \varphi^4 + O(\tilde \varphi^6)
  ,
\end{equation}
and see how it evolves  towards the limit $k\rightarrow 0$ according to the flow equation to obtain the physical effective potential $V_{\rm eff}^{\rm glue}:=V_{\rm eff,k=0}^{\rm glue}$.

As in (\ref{V_g}), $V_{\rm eff,k}^{\rm glue}$ is decomposed into two pieces:
\begin{equation}
 V_{\rm eff,k}^{\rm glue}  =   \hat V_{T,\hat k} +  \Delta \hat V_{\hat k} 
  ,
  \label{Veff}
\end{equation}
where we have defined the dimensionless potential  according to the rescaling (\ref{rescaling}).
The first part $\hat V_{T,\hat k}$ is the ($k$-dependent) perturbative part (\ref{V_{T,k}}) obtained essentially by the one-loop calculation with the regulator function $R_k$ being included.  For this part, the closed analytical form can be obtained, see Appendix~\ref{app:coefficient}.
While the second part $\Delta \hat V_{\hat k}$ represents the non-pertubative part which is initially zero $\Delta \hat V_{\hat k}|_{k=\Lambda}=0$ and is generated in the evolution of the renormalization group.  This part is obtained only by solving the flow equation (\ref{flow-eq}) and its analytical form is not available (at this moment).

We expand $\hat V_{T,\hat k}$ in powers of $\tilde\varphi=\varphi-\pi$: 
\begin{equation}
 \hat V_{T,\hat k}
=  A_{0,k} +  \frac{A_{2,k}}{2} \tilde \varphi^2 + \frac{A_{4,k}}{4!} \tilde \varphi^4 + O(\tilde \varphi^6)
  ,
  \label{VTcoeff}
\end{equation}
where coefficients are drawn as functions of $k$ in Fig.~\ref{fig:A_{2,k}}, see Appendix~\ref{app:coefficient} for their closed analytical forms.

\begin{figure}[h]
\centering
\includegraphics[width=3.0in,height=1.5in]{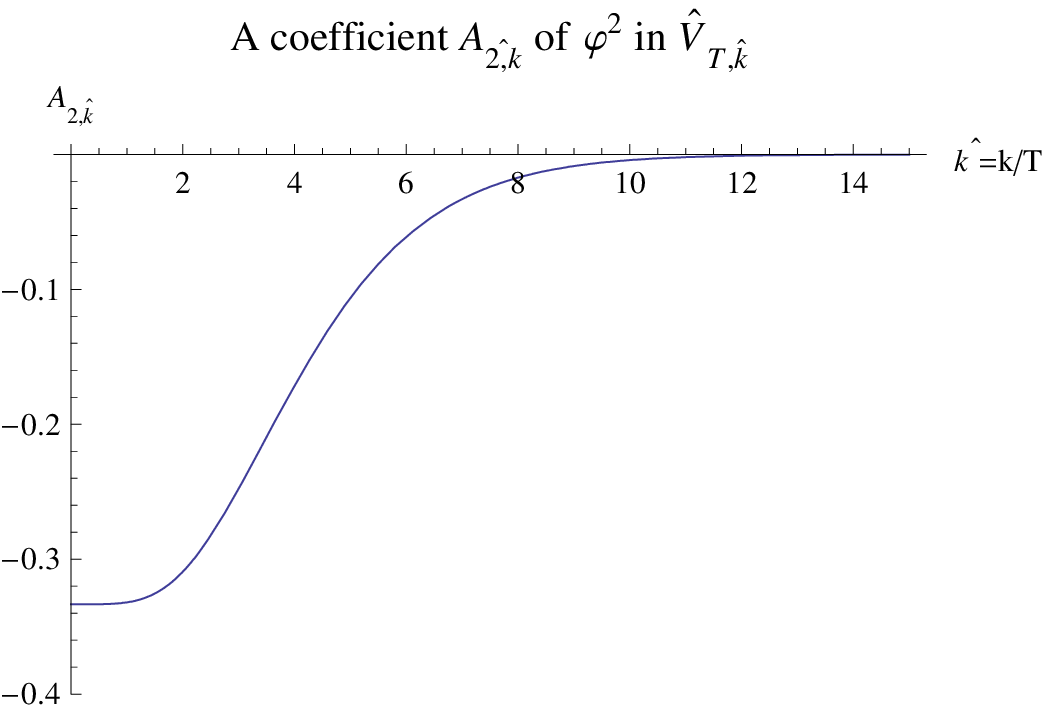}
\includegraphics[width=3.0in,height=1.5in]{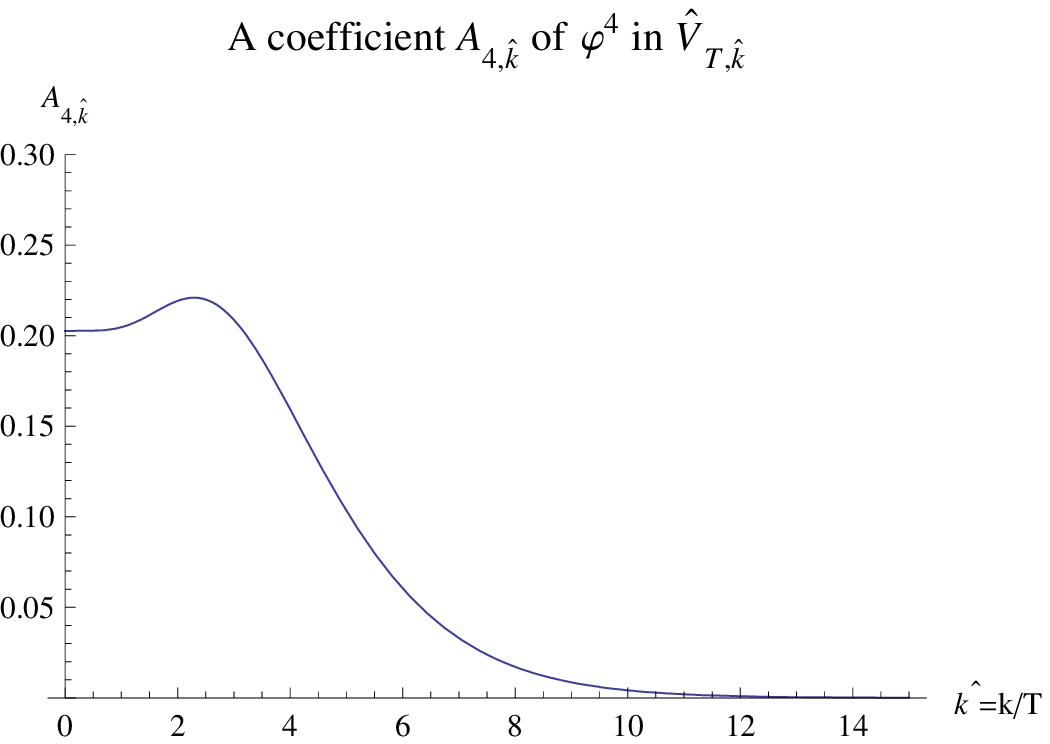}
\caption{$A_{2,k}$ and $A_{4,k}$ as functions of $\hat k$.}
\label{fig:A_{2,k}}
\end{figure}

Suppose that $\Delta \hat V_{\hat k}$ is of the form:
\begin{equation}
       \Delta \hat V_{\hat k}     
=    a_{0,k} +  \frac{a_{2,k}}{2} \tilde \varphi^2   + \frac{a_{4,k}}{4!} \tilde \varphi^4 + O(\tilde \varphi^6)
 .
 \label{V_T}
\end{equation}
A flow equation (\ref{flow-eq}) for the effective potential 
(\ref{V_g}) is reduced to a set of coupled flow equations for coefficients in the effective potential (\ref{Veff}) with (\ref{VTcoeff}) and (\ref{V_T}):
\begin{align}
    \partial_{\hat k}  a_{2,k}     
=&  -  \frac{(1+\frac15 \eta_k ) \hat k^2 }{6\pi^2} \frac{ \frac{4\pi\alpha_k}{\hat k^2}(A_{4,k}+a_{4,k})   }{[1+\frac{4\pi\alpha_k}{\hat k^2}   (A_{2,k}+a_{2,k})]^2} 
 ,
\nonumber\\
\partial_{\hat k}  a_{4,k}     
=& +   \frac{(1+\frac15 \eta_k ) \hat k^2 }{6\pi^2} \frac{6 [\frac{4\pi\alpha_k}{\hat k^2}(A_{4,k}+a_{4,k})]^2 }{[1+\frac{4\pi\alpha_k}{\hat k^2}   (A_{2,k}+a_{2,k})]^3}
 ,
\nonumber\\
\vdots  
\end{align}
which are coupled first-order ordinary but nonlinear differential equations for coefficients.
In Appendix~\ref{app:flow-equation}, we see  that this form (\ref{V_T}) is justified as a solution of the flow equation. In fact, it is easy to see  that  
$\partial_{\hat k} a_{1,k} =0$ and
$\partial_{\hat k} a_{3,k}=0$ are guaranteed from the flow equation, if the effective potential has no odd terms  at arbitrary $k$.
Therefore, if an initial condition, $a_{1,k} =0=a_{3,k}$ at $k=\Lambda$ is imposed, then   
$a_{1,k} \equiv 0$ and $a_{3,k} \equiv 0$ 
are maintained for $0 \le k \le \Lambda$ by solving the flow equation.
In performing numerical calculations, however,  one must truncate the infinite series of differential equations up to some finite order to obtain manageable set of equations.

\begin{figure}[h]
\centering
\includegraphics[width=3.0in,height=1.5in]{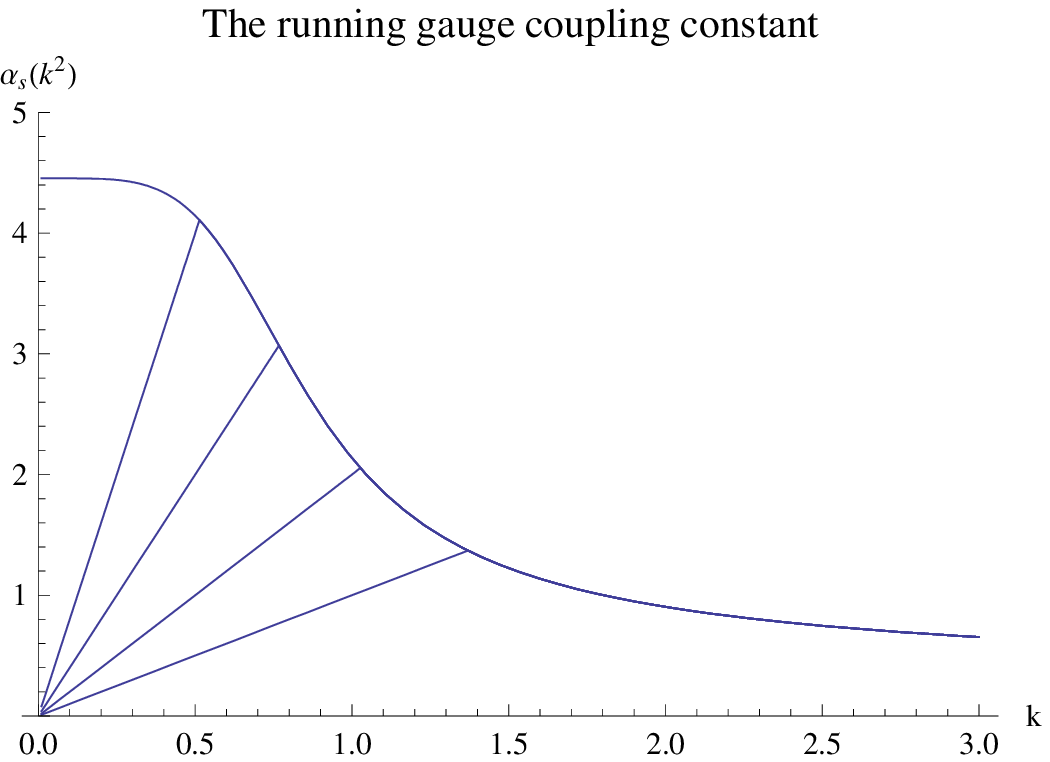}
\includegraphics[width=3.0in,height=1.5in]{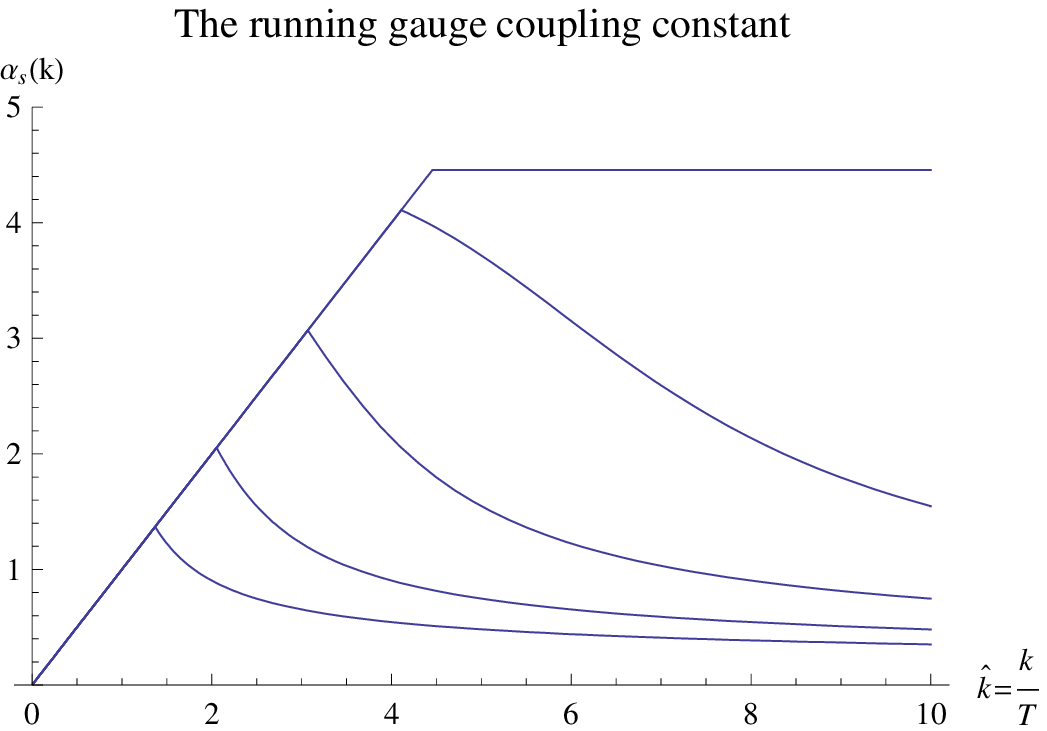}
\caption{The running gauge coupling constant $\alpha_k$ 
at $T=0.001, 0.125, 0.25, 0.50, 1.0$ GeV, from the top at the lowest temperature $T=0.001$GeV to the bottom  at the highest temperature $T=1.0$ GeV.
(The upper panel) $\alpha_k$ as functions of $k$.
(The lower panel) $\alpha_k$ as functions of $\hat{k}$.
For a given temperature, there is a critical value $\hat k_c$ separating the deep IR region (\ref{coupling-2}) from the higher momentum region (\ref{coupling-1}).
The discontinuity of the derivative seen at  $\hat k_c$ comes from a crude approximation in which we have taken into account just the first linear term (i.e., $c_1=c_2=\cdots=0$) in the expansion (\ref{coupling-2}), and can be avoided if we take into account higher order terms as explained below (\ref{coupling-2}). However, this is not essential to see qualitative behaviors of the solution of the flow equation.
}
\label{fig:alpha_k}
\end{figure}
\begin{figure}[h]
\centering
\includegraphics[width=3.0in,height=1.5in]{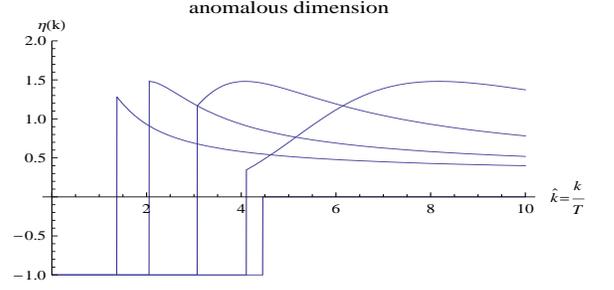}
\caption{The anomalous dimension $\eta_k$  as functions of $\hat k$
at $T=0.001, 0.125, 0.25, 0.50, 1.0$ GeV.
In each graph for a given temperature, there is a critical value $\hat k_c$ of $\hat k$ separating the deep IR region $\eta_k \simeq -1$ from the higher momentum (intermediate and UV) region $\eta_k >0$.
The temperature is distinguished by  $\hat k_c$ ranging from  the smallest value at the highest temperature $T=1.0$GeV to the largest value at the lowest temperature $T=0.001$ GeV where $\eta_k \simeq -1$ for $\hat k<\hat k_c$ and $\eta_k \simeq 0$ for $\hat k>\hat k_c$.
The discontinuity of the derivative seen at  $\hat k_c$ is due to the same reason as that explained in  Fig.~\ref{fig:alpha_k} and  is not essential to see qualitative behaviors of the solution of the flow equation.
}
\label{fig:eta_k}
\end{figure}

We can understand qualitatively why a 2nd order phase transition from the deconfinement phase to the confinement phase can occur by lowering the temperature.

\begin{figure}[h]
\centering
\includegraphics[width=2.5in,height=1.5in]{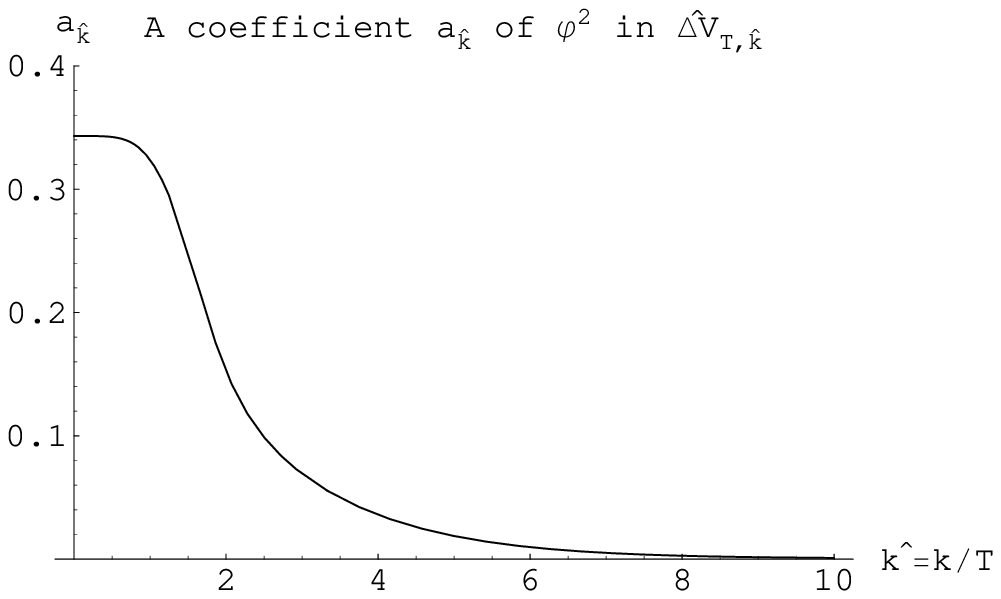}
\includegraphics[width=2.5in,height=1.5in]{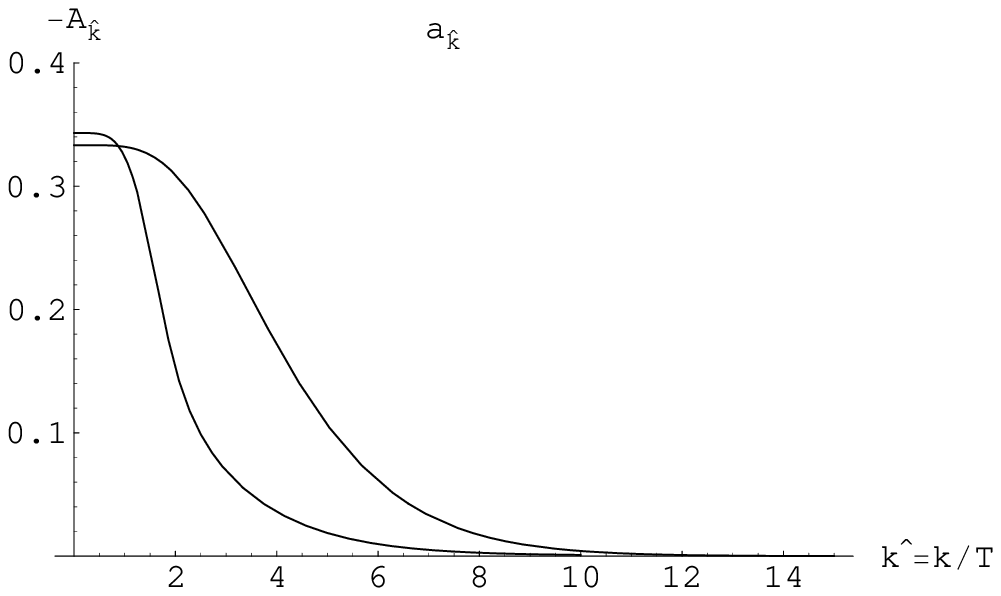}
\caption{$a_{2,k}$ vs. $-A_{2,k}$ as functions of $\hat k$ for $T<T_d$.}
\label{fig:mA_k}
\end{figure}

The flow starts from $a_{2,k}=0$ and hence $C_{2,k}=A_{2,k}+a_{2,k}<0$ (because of $A_{2,k}<0$) at $k=\Lambda \gg 1$.
We assume $C_{4,k}=A_{4,k}+a_{4,k}>0$ for $0 \le k \le \Lambda$, as a necessary condition for realizing a 2nd order transition. Otherwise, we must consider the higher-order terms, e.g. $O(\varphi^6)$.
 (This assumption is assured to be true by  numerical calculations of the full effective potential \cite{MP08,BGP07}, as reproduced in the previous section.)   
This assumption allows us to analyze just one differential equation for obtaining qualitative understanding:
\begin{align}
    \partial_{\hat k}  a_{2,k}     
=&  -  \frac{(1+\frac15 \eta_k )  }{6\pi^2} \frac{ 4\pi\alpha_k(A_{4,k}+a_{4,k})   }{[1+\frac{4\pi\alpha_k}{\hat k^2}   (A_{2,k}+a_{2,k})]^2} 
 .
 \label{flow-2}
\end{align}
Then the right-hand side of (\ref{flow-2}) is negative, since the running coupling constant $\alpha_k$ is positive and $1+\frac15 \eta_k$ is positive, see Fig.~\ref{fig:alpha_k} and Fig.~\ref{fig:eta_k}. 
Consequently, $a_{2,k}$ started at zero becomes positive $a_{2,k}>0$ just below $\Lambda$ and increases (monotonically) as $k$ decreases.
See Fig.~\ref{fig:mA_k}.

Note that the denominator can vanish  
$1+\frac{4\pi\alpha_k}{\hat k^2}   (A_{2,k}+a_{2,k})=0$ at some $k^*$ (since $C_{2,k}=A_{2,k}+a_{2,k}<0$ or $0<a_{2,k}<-A_{2,k}$) where the right-hand side of (\ref{flow-2}) becomes negative infinity and $a_{2,k}$ blows up there. 
To avoid this pathology and to obtain the solution all the way down to the limit $k \rightarrow 0$, $a_{2,k}$ must grow relatively rapidly so that $|A_{2,k}+a_{2,k}| \ll 1$ towards the limit $k \rightarrow 0$.

An important observation of the flow equation (\ref{flow-2}) is that  the explicit temperature-dependence comes from the running coupling constant alone.
At zero temperature, the running coupling constant is well parameterized by the fitting function \cite{FA02}:
\begin{equation}
     \alpha_k=\frac{4\pi \times 0.709/N_c}{\ln [e+a_1 (k^2)^{a_2}+b_1 (k^2)^{b_2}]}
 ,
 \label{coupling-0}
\end{equation}
where  
$
     a_1 = 5.292, \ a_2= 2.324, \ b_1 = 0.034, \ b_2 = 3.169 
 .
$
in units of GeV.
 
For the perturbative region $k \gg T$, i.e., $\hat k \gg 1$, we adopt this form:
$
k^2=T^2 \hat k^2  
$,
\begin{equation}
     \alpha_k = \frac{g_k^2}{4\pi} =\frac{4\pi \times 0.709/N_c}{\ln[e+a_1 (T^2 \hat k^2)^{a_2}+b_1 (T^2 \hat k^2)^{b_2}]}
 .
 \label{coupling-1}
\end{equation}
For the nonperturbative region $k < 2\pi T$, i.e., $\hat k < O(1)$, we adopt the running coupling which is governed by an infrared fixed point \cite{BG05}
\begin{align}
     \alpha_k =&  \alpha_{3d}^* \frac{k}{T} + c_1 \left( \frac{k}{T} \right)^2 + c_2 \left( \frac{k}{T} \right)^3 +  \cdots
\nonumber\\ 
=& \alpha_{3d}^* \hat k + c_1 \hat k^2 + c_2 \hat k^3 +  \cdots 
 ,
 \label{coupling-2}
\end{align}
where  coefficients $c_1, c_2,...$ are determined such that the coupling at zero temperature   (\ref{coupling-1})  and its derivative with respect to $k$ are connected continuously with this ansatz (\ref{coupling-2}) at the scale set by the lowest non-vanishing bosonic  Matsubara-mode $\omega=2\pi T$.
However, the running coupling constant at small momenta (\ref{coupling-2}) does not contribute to the explicit $T$-dependence in the scaled flow equation, since it is written in terms of the scaled $\hat{k}$ alone and hence denoted by a common curve going through the origin for any temperature $T$ in the second figure of Fig.~\ref{fig:alpha_k}. 
Therefore, the running coupling at very small momentum region can not be responsible for the confinement/deconfinement transition at finite temperature, if this observation is correct. 
As can be seen from the second figure of Fig.~\ref{fig:alpha_k}, the dominant contribution comes from the intermediate momentum region above $O(1)$ GeV. 
Thus, we can avoid the issue of gauge-fixing artifact in the deep IR region due to Gribov copies in the zero-temperature case, see e.g. \cite{FMP08,Kondo09} and reference therein.

We consider a solution of the reduced (or normalized) flow equation as a function of $\hat k$, rather than $k$, for a given temperature $T$. 
Then  the difference between high and low temperature phases attributes to the behavior of the running coupling constant $\alpha_k$  as a function of $k=T\hat k$, which brings the explicit $T$ dependence to the reduced flow equation. 
In the case of  high-temperature $T \gg 1$, 
$k=T\hat k$ becomes large for a wide range of $\hat k$ and 
the running coupling constant $\alpha_k$ remains relatively small.  The resulting slow increase of $a_{2,k}$ keeps $a_{2,k}$ small  such  that $C_{2,k}=A_{2,k}+a_{2,k}<0$ or $a_{2,k} <-A_{2,k}$ even at $k=0$. This leads to the center symmetry breaking at high-temperature.

In the case of low temperature  $T \ll1 $, 
$k=T\hat k$ becomes small  for the same range of $\hat k$ and  
the running coupling constant $\alpha_k$ gets into the intermediate region of O(1) GeV rapidly and becomes larger as the temperature becomes smaller.  
At sufficiently low temperature, $a_{2,k}$ increases in decreasing $\hat k$ so rapidly  that  $a_{2,k}$ eventually reaches to the point $A_{2,k}+a_{2,k}=0$ or $a_{2,k}=-A_{2,k}$ at a certain value $\hat k=\hat k_0$.  In other words, the graph of $a_{2,k}$ intersects with that of $-A_{2,k}$ at  $\hat k=\hat k_0$.
In the region $0<k<k_0$ where $C_{2,k}=A_{2,k}+a_{2,k}>0$ or $a_{2,k}>-A_{2,k}$, the flow equation reads
\begin{align}
    \partial_{\hat k}  a_{2,k}     
\simeq  -  \frac{(1+\frac15 \eta_k ) }{6\pi^2} \frac{ (A_{4,k}+a_{4,k}) \hat k^4  }{4\pi\alpha_k   (A_{2,k}+a_{2,k})^2} 
 ,
\end{align}
the right-hand-side gets small negative, and $a_{2,k}$ becomes flat near the IR limit. See Fig.~\ref{fig:mA_k}.
Finally,  $a_{2,k}$ reaches the value realizing  $C_{2,k}=A_{2,k}+a_{2,k}>0$ or $a_{2,k}>-A_{2,k}$ at $k=0$. 
This leads to the recovery of the center symmetry.
The difference is clearly seen from the second figure of Fig.~\ref{fig:alpha_k} where the running gauge coupling $\alpha_k$ is drawn as a function of $\hat k$ for various temperatures. 

In our treatment, the difference between the three-dimensional RG scale $k_T$ and the four-dimensional one $k$ is neglected by equating two scales $k_T=k$ just for simplifying the analysis, since it is enough for obtaining a qualitative understanding for the transition. This is not be the case for obtaining quantitative results, see Appendix C of \cite{MP08} for the precise treatment on this issue.

\section{Quark part and gauged nonlocal NJL model}\label{sec:NJL}

We examine the quark self-interaction part
$S_{\rm int} =   \int d^Dx \int d^Dy  \frac{1}{2}  \mathscr{J}^{\mu A}(x)  g^2(Q^{-1}[\mathscr{V}])_{\mu\nu}^{AB}(x,y) \mathscr{J}^{\nu B}(y)
$.
In  estimating the effect of $Q^{-1}[\mathscr{V}]$, we take the  same  approximation as the above. 
Consequently, the inverse  $(Q^{-1})_{\mu\nu}^{AB}[\mathscr{V}]$  is diagonal in the Lorentz indices:
\begin{align} 
& (Q^{-1})_{\mu\nu}^{AB}[\mathscr{V}]  
= g^{\mu\nu} (G^{-1})^{AB}[\mathscr{V}]  
\nonumber\\
=&     g^{\mu\nu}
  \begin{pmatrix}
  \frac12 [F_{\varphi}+F_{-\varphi}] & -\frac{1}{2i} [F_{\varphi}-F_{-\varphi}] & 0 \cr
  \frac{1}{2i} [F_{\varphi}-F_{-\varphi}] & \frac12 [F_{\varphi}+F_{-\varphi}] & 0 \cr
  0 & 0 & F_{0} \cr
  \end{pmatrix} 
 ,
\end{align}
where $F_{\varphi}$ is defined by
\begin{align} 
F_{\varphi}(i\partial) :=&  \frac{1}{(i\partial_\ell)^2+(i\partial_0+T\varphi)^2}
\nonumber\\ 
=&  \frac{1}{(i\partial_\mu)^2+(T\varphi)^2+2T\varphi i\partial_0}
 . 
\end{align}
In what follows, we consider only the diagonal parts of $(G^{-1})^{AB}[\mathscr{V}]$.  
This is achieved by the procedure
\begin{equation}
  \frac{g^2}{2} (Q^{-1})_{\mu\nu}^{AB}(x,y) = g^{\mu\nu}   \delta^{AB} \mathcal{G}(x-y)
  ,
\end{equation}
which yields
\begin{equation}
 \mathcal{G}(x-y)  = \frac{g^2}{2} (Q^{-1})_{\mu\nu}^{AB}(x,y)  \frac{g_{\mu\nu}}{D}  \frac{\delta^{AB}}{N_c^2-1} 
  .
\end{equation}
Then the nonlocal interaction is obtained as
\begin{equation}
 \mathcal{G}(x-y) 
=  \frac{g^2}{2}  \frac{{\rm tr}(G^{-1})}{N_c^2-1} 
=  \frac{g^2}{2} \frac{F_{\varphi}+F_{-\varphi}+F_{0}}{3}
  .
\end{equation}
This approximation is used just for simplifying the Fierz transformation performed below and hence it can be improved by taking into account the off-diagonal parts of $G^{-1}$ if it is necessary to do so.

For $D=4$,  we use the Fierz identity \cite{Fierz} to rewrite the nonlocal current-current interaction as
\begin{align}
   S_{\rm int}   
   =& \int d^4x \int d^4y      \mathscr{J}^{\mu A}(x) \mathcal{G}(x-y) \mathscr{J}^{\mu A}(y)
  \nonumber\\
=&    \int d^4x  \int d^4y   \mathcal{G}(x-y) \sum_{\alpha} c_\alpha (\bar{\psi}(x) \Gamma_\alpha   \psi(y) ) 
\nonumber\\& 
 \times (\bar{\psi}(y) \Gamma_\alpha  \psi(x) ) 
  \nonumber\\
   =&   \int d^4x   \int d^4z  \mathcal{G}(z) \sum_{\alpha} c_\alpha \{ \bar{\psi}(x+z/2) \Gamma_\alpha   \psi(x-z/2) \}   
  \nonumber\\&
\times  \{ \bar{\psi}(x-z/2) \Gamma_\alpha  \psi(x+z/2) \} ,
\end{align}
where the $\Gamma_\alpha$ are a set of Dirac spinor, color and flavor matrices, resulting from the Fierz transform, with the property $\gamma_0 \Gamma_\alpha^\dagger \gamma_0=\Gamma_\alpha$. 
Although the Fierz transformation induces mixings and recombinations among operators, the resulting theory must maintain the symmetries of the original QCD Lagrangian. 
A minimal subset of operators satisfying the global chiral symmetry $SU(2)_L \times SU(2)_R$ which governs low-energy QCD with two-flavors  is the color-singlet of scalar-isoscalar and pseudoscalar-isovector operators 
Thus, by restricting $\Gamma_\alpha$ hereafter to  
\begin{equation}
\Gamma_\alpha:=(\mathbf{1}, i\gamma_5 \vec{\tau})
\end{equation}
and ignoring other less relevant operators (vector and axial-vector terms in color singlet and color octet channels), 
 we arrive at a nonlocal gauged NJL model   
\begin{align}
   S_{\rm eff}^{\rm gNJL} =&  \int d^4x \bar\psi(x) ( i\gamma^\mu \mathcal{D}_\mu[\mathscr{V}] - \hat m_q+i\gamma^0 \mu_q) \psi(x) + S_{\rm int} ,
 \nonumber\\
  S_{\rm int} =&  \int d^4x \int  d^4z \mathcal{G}(z) [\bar \psi(x+z/2) \Gamma_\alpha \psi(x-z/2)  
 \nonumber\\&
\times  \bar \psi(x-z/2)  \Gamma_\alpha \psi(x+z/2)    ] 
 .
 \label{gNJL}
\end{align}
This form is regarded as a gauged version of the nonlocal NJL model proposed in \cite{HRCW08}.    
The function $\mathcal{G}(z)$ is replaced by a coupling constant $G$ times a normalized distribution $\mathcal{C}(z)$:
\begin{equation}
 \mathcal{G}(z)  :=  \frac{G}{2} \mathcal{C}(z) ,
 \quad
 \int d^4z \mathcal{C}(z) = 1 
 .
\end{equation}
The standard (local) gauged NJL model follows for the limiting case $\mathcal{C}(z)=\delta^4(z)$ 
with
$\int d^4z \mathcal{C}(z)=1$.

In contrast to \cite{HRCW08}, however, $G$ and $\mathcal{C}$ are determined in conjunction with the behavior of the Polyakov loop $L$ or $\varphi$ at temperature $T$:
using the Fourier transform $\tilde{\mathcal{G}}(p)$ of $\mathcal{G}$, they are expressed as
\begin{equation}
   \frac{G}{2}  
=   \tilde{\mathcal{G}}(p=0) ,
   \quad
   \tilde{\mathcal{C}}(p) = \tilde{\mathcal{G}}(p)/\tilde{\mathcal{G}}(p=0)
  ,
  \label{G}
\end{equation}
where 
\begin{align}
 \tilde{\mathcal{G}}(p)
=&   \frac{g^2}{2}  \frac{\tilde{F}_{\varphi}(p)+\tilde{F}_{-\varphi}(p)+\tilde{F}_{0}(p)}{3} , 
\\
\tilde{F}_{\varphi}(p)=& \frac{1}{p^2+(T\varphi)^2+2T\varphi p_0} .
\nonumber
\end{align}
Note that $\tilde{F}_{\varphi}(p=0)$ and hence $G$ diverge at $T=0$.
This comes from an improper treatment of the $T=0$ part.
To avoid  this IR divergence at $T=0$, we add the $T=0$ contribution $M_0^2 \simeq M_X^2$ and replace $F_{\varphi}(i\partial)$ by  
\begin{align} 
F_{\varphi}(i\partial) 
&= \frac{1}{(i\partial_\ell)^2+(i\partial_0+T\varphi)^2+M_0^2}
\nonumber\\
=& \frac{1}{(i\partial_\mu)^2+(T\varphi)^2+2T\varphi i\partial_0+M_0^2}
 , 
\end{align}
and
\begin{align}
\tilde{F}_{\varphi}(p)
&= \frac{1}{\bm{p}^2+(p_0+T\varphi)^2+M_0^2}
\nonumber\\
=& \frac{1}{p^2+(T\varphi)^2+2T\varphi p_0+M_0^2} .
\end{align}
In fact, such a contribution $\frac12 M_0^2 $ comes in $G^{AB}$ as an additional term $M_0^2 \delta^{AB}$ from the $O(\mathscr{X}^4)$ terms (Note that $O(\mathscr{X}^3)$ terms are absent for $G=SU(2)$), as already mentioned in the above. 

Another way to avoid this IR divergence is to introduce the regulator term which is needed to improve the one-loop perturbative result and obtain a nonperturbative one according to the Wilsonian renormalization group:
\begin{align}
 \Delta S_k =& \int d^Dx \frac12 \mathscr{X}_\mu^A(x) [\delta^{AB} g_{\mu\nu} R_k(i\partial)] \mathscr{X}_\nu^B(x) 
 \nonumber\\
 =& \int \frac{d^Dp}{(2\pi)^D} \frac12 \tilde{\mathscr{X}}_\mu^A(-p) [\delta^{AB} g_{\mu\nu} \tilde{R}_k(p)] \tilde{\mathscr{X}}_\nu^B(p) 
  ,
\end{align}
where $k$ is the RG scale and $\tilde{R}_k(p)$ is the Fourier transform of $R_k(i\partial)$. 
The regulator function $R_k$ introduces a mass proportional to $k^2$, which plays a similar role to $M_0^2$ in the above, as long as $k >0$.

The NJL model \cite{NJL61} is well known as a low-energy effective theory of QCD to describe the dynamical breaking of chiral symmetry in QCD  (at least in the confinement phase), see e.g. \cite{Klevansky92,HK94}. 
The theory given above by $S_{\rm eff}^{\rm QCD}=S_{\rm eff}^{\rm glue}+S_{\rm eff}^{\rm gNJL}$ is able to describe chiral-symmetry breaking/restoration and quark confinement/deconfinement on an equal footing where 
the pure gluon part $S_{\rm eff}^{\rm glue}$  describes confinement/deconfinement transition signaled by the Polyakov loop average. 
We can incorporate the information on confinement/deconfinement transition into the quark sector through the covariant derivative $\mathcal{D}[\mathscr{V}]$ and the nonlocal NJL interaction $\mathcal{G}$ ($G$ and $\mathcal{C}$), in sharp contrast to the conventional PNJL model where the entanglement between chiral-symmetry breaking/restoration and confinement/deconfinement was incorporated  through the covariant derivative $\mathcal{D}[\mathscr{V}]$ alone and the nonlocal NJL interaction $\mathcal{G}$ is fixed to the zero-temperature case. 
In our theory, the nonlocal NJL interaction $\mathcal{G}$ ($G$ and $\mathcal{C}$) is automatically determined through the information of confinement/deconfinement dictated by the Polyakov loop $L$ (non-trivial gluon background), while  in the nonlocal PNJL model \cite{HRCW08} the low-momentum (non-perturbative) behavior of $\mathcal{C}$ was not controlled by first principles and was provided by the instanton model.

To study chiral dynamics, it is convenient to bosonize the gauged nonlocal NJL model as done \cite{HRCW08}.
The nonlocal gauged NJL model (\ref{gNJL}) can be bosonised as follows. 
Define
\begin{equation}
\Phi_\alpha(x) :=(\sigma(x), \vec\pi(x)) .
\end{equation}
To eliminate the quadratic term in the nonlocal currents, 
we insert the unity:
\begin{align}
 1=& \int \mathcal{D} \sigma \mathcal{D} \vec{\pi} \exp \Big\{ -\int d^4z \mathcal{C}(z) 
\nonumber\\&
 \times \int d^4x \frac{1}{2G} [\Phi_\alpha(x) + G\bar\psi(x+z/2) \Gamma_\alpha \psi(x-z/2)]
\nonumber\\&
 \times [\Phi_\alpha(x) + G\bar\psi(x+z/2) \Gamma_\alpha \psi(x-z/2)]^* \Big\} 
  ,
\end{align}
where we have used
$\int d^4z \mathcal{C}(z)=1 $.
Then we have the gauged Yukawa model: 
\begin{align}
&  \int \mathcal{D} \bar\psi  \mathcal{D} \psi e^{-S_{\rm eff}^{\rm gNJL} }
\nonumber\\
 =&  
 \int \mathcal{D} \bar\psi  \mathcal{D} \psi 
\int \mathcal{D} \sigma \mathcal{D} \vec{\pi} 
\exp  \{ - S_{\rm eff}^{\rm gY}
 \} 
  ,
\end{align}
where with $x^\prime:=x+z/2$, $y^\prime:=x-z/2$, 
\begin{align}
   S_{\rm eff}^{\rm gY} 
 =&  
  \int d^4x^\prime \int d^4y^\prime  \bar\psi(x^\prime) \Big[ \nonumber\\
& \delta^4(x^\prime-y^\prime) 
(-i\gamma^\mu \mathcal{D}_\mu[\mathscr{V}] + \hat m_q+i\gamma^4 \mu_q)
\nonumber\\
&+ \frac12 \mathcal{C}(x^\prime-y^\prime) \Gamma_\alpha [\Phi_\alpha(\frac{x^\prime+y^\prime}{2}) + \Phi_\alpha^*(\frac{x^\prime+y^\prime}{2})] 
\Big] \psi(y^\prime)
\nonumber\\
&+ \int d^4x \frac{1}{2G}  \Phi_\alpha(x)\Phi_\alpha^*(x)  
  ,
\end{align}
or
\begin{align}
&   S_{\rm eff}^{\rm gY} 
\nonumber\\
 =&  
 \int \frac{d^4p}{(2\pi)^4} \frac{d^4p^\prime}{(2\pi)^4}  \bar\psi(p) \Big[ 
\nonumber\\
& (2\pi)^4 \delta^4(p-p^\prime) (- \gamma^\mu (p_\mu + g \tilde{\mathscr{V}}_\mu(p)) + \hat m_q+i\gamma^4 \mu_q)
\nonumber\\
&+ \frac12 \tilde{\mathcal{C}}(\frac{p+p^\prime}{2}) \Gamma_\alpha [\Phi_\alpha(p-p^\prime) + \Phi_\alpha^*(p-p^\prime)]  \Big] \psi(p^\prime)
\nonumber\\&
 + \int \frac{d^4p}{(2\pi)^4} \frac{1}{2G}  \Phi_\alpha(p)\Phi_\alpha^*(p)
   .
\end{align}

Finally, the bosonized theory of the gauged NJL model is obtained by way of the gauged Yukawa model by integrating out quark fields as
\begin{align}
&  \int \mathcal{D} \bar\psi  \mathcal{D} \psi e^{-S_{\rm eff}^{\rm gNJL} }
\nonumber\\
 =&  
\int \mathcal{D} \sigma \mathcal{D} \vec{\pi} 
\int \mathcal{D} \bar\psi  \mathcal{D} \psi 
\exp  \{ -S_{\rm eff}^{\rm gY}
 \} 
\nonumber\\
 =& \int \mathcal{D} \sigma \mathcal{D} \vec{\pi}  \exp \{ - S_{\rm eff}^{\rm boson} \}
  ,
\end{align}
where the bosonised action $S_{\rm eff}^{\rm boson}$ is 
\begin{align}
   S_{\rm eff}^{\rm boson}
 =&  
 - {\rm Tr} \ln  \Big\{ 
 \delta^4(x^\prime-y^\prime) 
(-i\gamma^\mu (\partial_\mu -ig \mathscr{V}_\mu) +i\gamma^4 \mu_q)
\nonumber\\
&+ \hat m_q + \frac12 \mathcal{C}(x^\prime-y^\prime) \Gamma_\alpha [\Phi_\alpha(\frac{x^\prime+y^\prime}{2}) + \Phi_\alpha^*(\frac{x^\prime+y^\prime}{2})] \Big\} 
\nonumber\\
&+ \int d^4x \frac{1}{2G}  \Phi_\alpha(x)\Phi_\alpha^*(x)  
  ,
\end{align}
or
\begin{align}
 S_{\rm eff}^{\rm boson}
 =&  
 - {\rm Tr} \ln  \Big\{ 
(2\pi)^4 \delta^4(p-p^\prime)  [- \gamma^\mu (p_\mu + g \tilde{\mathscr{V}}_\mu) +i\gamma^4 \mu_q]
\nonumber\\&
+ \hat m_q + \frac12 \tilde{\mathcal{C}}(\frac{p+p^\prime}{2}) \Gamma_\alpha [\Phi_\alpha(p-p^\prime) + \Phi_\alpha^*(p-p^\prime)]  \Big\}  
\nonumber\\&
+ \int \frac{d^4p}{(2\pi)^4} \frac{1}{2G}  \Phi_\alpha(p)\Phi_\alpha^*(p)
  .
\end{align}

\section{Implications of the Polyakov loop for chiral-symmetry breaking at finite temperature}\label{sec:}

The thermodynamics of QCD can be studied based on our effective theory derived in this paper in the similar way to the nonlocal PNJL model \cite{HRCW08}. 
But this must be done by including the effect of gluon properly. 
In the PNJL model, the effect of the gluon was introduced by the standard minimal gauge coupling procedure, i.e., replacing the normal derivative $\partial_\mu$ by the covariant derivative $D_\mu[\mathscr{A}]:=\partial_\mu -ig\mathscr{A}_\mu$.  
In our effective theory, the effect of the gluon is introduced through the NJL coupling constant $G$ and the nonlocality function $\mathcal{C}$, in addition to the minimal coupling $D_\mu[\mathscr{V}]$.  
The nonlocal NJL interaction among quarks are mediated by gluons at finite temperature in QCD. 
Therefore, both $G$ and  $\mathcal{C}$ characterizing nonlocal NJL interaction inevitably have temperature dependence, which could be different depending on whether quarks are in confinement or deconfinement phases. 
 
At $T=0$, QCD must be in the hadron phase where the chiral symmetry is spontaneously broken, which means that the NJL coupling constant $G(0)$ at zero temperature must be greater than the critical NJL coupling constant $G_c$:
\begin{equation}
 G(0) =  g^2  \frac{1}{M_0^2} > G_c .
\end{equation}
The nonlocality function or the form factor $\tilde{\mathcal{C}}(p)$ at $T=0$ behaves
\begin{equation}
 \tilde{\mathcal{G}}(p)
=   \frac{g^2}{2}     \frac{1}{p^2 +M_0^2 }, \ 
 \tilde{\mathcal{G}}(0)
=   \frac{g^2}{2}    \frac{1}{M_0^2 } ,
\end{equation}
and
\begin{equation}
\tilde{\mathcal{C}}(p) =  \frac{M_0^2}{p^2 +M_0^2 } 
    .
\end{equation}

As an immediate outcome of our effective theory, this determines the temperature-dependence of the coupling constant $G$ of nonlocal NJL model.
Using (\ref{G}), we have
\begin{equation}
 G(T) =   \frac13 g^2 \left[ \frac{2}{(T\varphi)^2+M_0^2}+\frac{1}{M_0^2} \right] , \
\end{equation}
which lead to the NJL coupling constant normalized at $T=0$:
\begin{equation}
 G(T)/G(0) =   \frac13  \left[ \frac{2M_0^2}{(T\varphi)^2+M_0^2}+1 \right] .
\end{equation}

In the presence of the dynamical quark $m_q < \infty$, the Polyakov loop is not an exact order parameter and does not show a sharp charge with discontinuous derivatives.  Even in this case, we can introduce the pseudo critical temperature $T_d^*$ as a temperature achieving the peak of the susceptibility. 
Below the deconfinement temperature $T_d^*$, i.e., $T < T_d^*$, therefore, $L \simeq 0$ or $\varphi \simeq \pi$, the NJL coupling constant $G$ has the temperature-dependence
\begin{equation}
G(T)/G(0) \simeq   \frac13  \left[ \frac{2M_0^2}{\pi^2 T^2+M_0^2}+1 \right]  \ (T < T_d^*) .
\end{equation}

This naive estimation gives a qualitative understanding for the existence of chiral phase transition.   Since $G(T)$ is (monotonically) decreasing as the temperature $T$ increases, it becomes smaller than the critical NJL coupling constant  
\begin{equation}
 G(0) =  g^2  \frac{1}{M_0^2} > G_c ,
 \quad 
  T \uparrow \infty \Longrightarrow 
G \downarrow 0 .
\end{equation}
Thus, the chiral transition temperature $T_\chi$ will be determined (if the chiral-symmetry restoration and confinement coexist or the chiral symmetry is restored in the confinement environment before deconfinement takes place, i.e.,  $T_\chi \le T_d^*$ ) by solving
\begin{equation}
 G(T_\chi) \equiv   \frac{G(0)}{3}  \left[ \frac{2M_0^2}{T_\chi^2 \pi^2 +M_0^2}+ 1\right] = G_c 
 .
\end{equation}
Here we have assumed that the nonlocality function $\tilde{\mathcal{C}}(p)$ gives the dominant contribution at $p=0$, namely, $\tilde{\mathcal{C}}(p) \le  \tilde{\mathcal{C}}(0)=\int d^4z \mathcal{C}(z)=1$ and that the occurrence of the chiral transition is determined by the NJL coupling constant alone. 

At finite temperature $T$, the form factor reads
\begin{align}
& \mathcal{C}(\bm{x}-\bm{y}) 
\nonumber\\ 
 =& T \sum_{n \in \mathbb{Z}} \int \frac{d^3p}{(2\pi)^3} 
  \tilde{\mathcal{C}}(p_0=\omega, \bm{p}) e^{i\bm{p} \cdot (\bm{x}-\bm{y})} 
\nonumber\\ 
 =& T \sum_{n \in \mathbb{Z}} \int \frac{d^3p}{(2\pi)^3} 
\frac{M_0^2}{3} \Big[ \frac{2}{\bm{p}^2+(\omega+T\varphi)^2+M_0^2} 
\nonumber\\&
+ \frac{1}{\bm{p}^2+\omega^2+M_0^2} \Big] e^{i\bm{p} \cdot (\bm{x}-\bm{y})} 
\nonumber\\ 
 =&   \int \frac{d^3p}{(2\pi)^3} 
\frac{M_0^2}{6\epsilon_p} \Big[ 2\frac{\sinh (\epsilon_p/T)}{\cosh(\epsilon_p/T)-\cos(\varphi)} 
\nonumber\\&
+  \frac{\sinh (\epsilon_p/T)}{\cosh(\epsilon_p/T)-1}  \Big] e^{i\bm{p} \cdot (\bm{x}-\bm{y})} 
 ,
\end{align}
where we have defined 
$\epsilon_p:=\sqrt{\bm{p}^2+M_0^2}$
and used
\begin{align}
 T \sum_{n \in \mathbb{Z}}  \frac{1}{(\omega+C)^2+\epsilon_p^2}   
 = \frac{1}{2\epsilon_p}   \frac{\sinh (\epsilon_p/T)}{\cosh(\epsilon_p/T)-\cos(C/T)} 
 .
\end{align}
The form factor $\mathcal{C}$ does not change so much around the deconfinement temperature $T \sim T_d^*$ (or $\varphi \sim \pi$). 
This is reasonable since the form factor is nearly equal to the $\mathscr{X}$ correlator $Q^{-1}$, as already  mentioned in sec. II. 

For more precise treatment, we must obtain the full effective potential $V_{\rm eff}(\sigma, \varphi)$ as a function of two order parameters $\sigma$ (or $\langle \bar\psi \psi \rangle$) and $\varphi$ (or $\langle L \rangle$), and look for a set of values $(\sigma, \varphi)= (\sigma_0, \varphi_0)$ at which the minimum $V_{\rm eff}(\sigma_0, \varphi_0)$ of $V_{\rm eff}(\sigma, \varphi)$ is realized. 
Then $\varphi$ must be replaced by $\varphi_0$ in the above consideration. 
For this goal, we must develop the RG treatment for the full theory. 
This issue will be studied in a subsequent paper. 

\section{How to understand the entanglement between confinement and chiral symmetry breaking}\label{sec:entanglement}

To discuss the entanglement between confinement and chiral symmetry breaking, we wish to obtain the total effective potential $V_{}^{\rm QCD}$ of QCD written in terms of two order parameters, i.e., the Polyakov loop average $\langle L \rangle$ and chiral condensate $\langle \bar \psi \psi \rangle$, so that its minima determine the vacuum for a given set of parameters $m_q$, $T$ and $\mu_q$ when $N_c$ and $N_f$ are fixed. 
Here $m_q \uparrow \infty$ is the pure Yang-Mills limit and $m_q \downarrow 0$ is the chiral limit.

The effective potential for the quark part is obtained by integrating out quark degrees of freedom.
The simplest form is obtained  e.g., from the bosonized model as
\begin{align}
   & V_{}^{\rm quark}
\nonumber\\
 =&  - {\rm Tr} \ln \Big\{  i\gamma^\mu \partial_\mu  
+ m_q + \mathcal{C} \sigma  
 -    g \mathscr{A}_4  \gamma^4  
 + i   \mu_q \gamma^4  \Big\} 
 + \frac{1}{2G}  \sigma^2 
  .
\label{V^q}
\end{align}
Then the RG-scale $k$ dependent effective potential $V_{k}^{\rm quark}$ for the quark part must be given as the solution of the flow equation.  
In the same approximation as the above, it is written in terms of two order parameters $\sigma$ and $\varphi$: 
\begin{align}
  &  V_{k}^{\rm quark}(\sigma,\varphi)_{m_q,T,\mu_q}
\nonumber\\ 
=&      
 -  T \sum_{n \in \mathbb{Z}} \int \frac{d^3p}{(2\pi)^3} {\rm tr} \ln \Big[   i \omega_n \gamma^0- p_j \gamma^j  + m_q +  \mathcal{C}(p)  \sigma
\nonumber\\&
\quad\quad\quad\quad \quad\quad\quad\quad\quad\quad
- T\varphi T_3 \gamma^4 + i  \mu_q \gamma^4 + R_k^{\rm quark} \Big]  
\nonumber\\&
 + \frac{1}{2G}  \sigma^2 
  ,
  \label{V^q_k}
\end{align}
where $T_3=\sigma_3/2$ and $R_k^{\rm quark}$ is the regulator function for quarks.
In the limit $k \downarrow 0$, indeed, $V_{k}^{\rm quark}$ (\ref{V^q_k}) reduces to $V_{}^{\rm quark}$ (\ref{V^q}). 
The effective potential $V_{k}^{\rm quark}$ (\ref{V^q_k}) depends on $m_q$, $T$ and $\mu_q$ when $Nc$ and $N_f$ are fixed. 
Due to the $p_0$ dependence of the ``mass'' function $M(p)$ which is an immediate consequence of the nonlocality of the present NJL model, it is difficult to obtain the closed analytical form by performing the summation over the Matsubara frequencies.

In our strategy, a full effective potential $V_{{\rm eff},k}^{\rm QCD}(\sigma,\varphi)$ of QCD is given by summing three parts:
\begin{equation} 
 V_{{\rm eff},k}^{\rm QCD}(\sigma,\varphi)
 = V_{k}^{\rm glue}(\varphi) + V_{k}^{\rm quark}(\sigma,\varphi) + \Delta V_{k}^{\rm QCD}(\sigma,\varphi)  
 ,
\end{equation}
with  the pure gluon part $V_{k}^{\rm glue}(\varphi) = V_{T,k}$ (\ref{V_g}), 
\begin{align}
 V_{k}^{\rm glue}(\varphi) 
=&    {\rm Tr}  \{  \ln [G^{AB}+\delta^{AB}R_{k} ] \} ,
\nonumber\\
 =& {\rm Tr}  \{  \ln [
   - \delta^{AB} \partial_\mu^2  +   (\delta^{AB}  - \delta^{A3}\delta^{B3})(T \varphi)^2 
\nonumber\\&
\quad\quad\quad\quad
+ 2  \epsilon^{AB3} T \varphi \partial_0 
+\delta^{AB}R_{k} ] \} ,
\end{align}
 the quark part (\ref{V^q_k}), 
\begin{align}
   V_{k}^{\rm quark}(\sigma,\varphi)
   =&    \frac{1}{2G}  \sigma^2
 - {\rm Tr} \ln \{ 
 i\gamma^\mu \partial_\mu  + m_q + \mathcal{C} \sigma  
- T\varphi T_3 \gamma^4 
\nonumber\\& 
\quad\quad\quad\quad\quad 
 + i  \mu_q \gamma^4 + R_k^{\rm quark}
\} ,
\end{align}
and  a non-perturbative part $\Delta V_{k}^{\rm QCD}(\sigma,\varphi)$ induced in the RG evolution according to a flow equation.
We assume that the total effective action of QCD obtained after integrating out the fields other than those relevant to chiral symmetry and confinement is the form
\begin{align}
 \Gamma_k  =& \int_{0}^{1/T}dx_4 \int d^3x  \Big\{ 
 \frac{1}{2}Z_0  [ \partial_j  \mathscr{V}_0(\bm{x})]^2
+ \frac{1}{2} Z_\sigma [\partial_j \sigma(x)]^2  
\nonumber\\&
+ V_{{\rm eff},k}^{\rm QCD}(\sigma,\varphi)  \Big\} 
  ,
\end{align}
and obeys the flow equation:
\begin{align}
\partial_t \Gamma_k 
=&  \frac12 {\rm Tr} \left\{  
\left[ \frac{\overrightarrow{\delta}}{\delta \sigma^\dagger} \Gamma_k \frac{\overleftarrow{\delta}}{\delta \sigma} + R_{k} \right]^{-1} \cdot 
\partial_t R_{k} \right\}
\nonumber\\
 &+ \frac12 {\rm Tr} \left\{ \left[ \frac{\overrightarrow{\delta}}{\delta \mathscr{V}^\dagger} \Gamma_k \frac{\overleftarrow{\delta}}{\delta \mathscr{V}} + R_{k} \right]^{-1} \cdot \partial_t R_{k} \right\} 
  .
\end{align}

If the flow equation was solved, we would have obtained the effective potential of QCD, $V_{{\rm eff},k}^{\rm QCD}(\sigma,\varphi)$ which has the following power-series expansion with respect to two variables $\sigma$ and $\tilde\varphi$ in the neighborhood of the transition point where $\sigma=0=L$ according to the Landau argument (as demonstrated in the pure glue case). 
\begin{align}
 V_{{\rm eff}}^{\rm QCD}(\sigma,\tilde\varphi)
 =& V^{\rm g}(\tilde\varphi)   + V^{\rm q}(\sigma)
+ V^{\rm c}(\sigma, \tilde\varphi)  ,
\nonumber\\
 V^{\rm g}(\tilde\varphi) =& C_0 + C_1 \tilde\varphi +  \frac{C_2}{2} \tilde\varphi^2 + \frac{C_3}{3} \tilde\varphi^3 + \frac{C_4}{4} \tilde\varphi^4 + O(\tilde\varphi^6)
 ,
\nonumber\\
  V^{\rm q}(\sigma) =&   \frac{E_2}{2} \sigma^2 + \frac{E_4}{4} \sigma^4 + O(\sigma^6)
   ,
\nonumber\\
  V^{\rm c}(\sigma, \tilde\varphi) =&  
    F_1 \sigma^2 \tilde\varphi  + \cdots 
 ,
\end{align}
where 
$V^{\rm g}(\hat\varphi)$ denotes a part written in terms of $\hat\varphi$ alone, 
and $V^{\rm q}(\sigma)$ denotes a part written in terms of $\sigma$ alone, 
while $V^{\rm c}(\sigma, \hat\varphi)$ denotes the cross term  between $\sigma$ and $\hat\varphi$.

Once dynamical quarks are introduced, the exact center symmetry in pure Yang-Mills theory is no longer intact. 
Therefore, the QCD effective potential includes the explicitly center-symmetry-breaking term.
For $G=SU(2)$,  the center symmetry $\tilde\varphi \rightarrow - \tilde\varphi$ is explicitly broken as  $C_1 \not=0$, $C_3 \not=0$ in $V^{\rm g}(\varphi)$ and $F_1 \not= 0$ in $V^{\rm c}(\sigma, \varphi)$. 
The existence of the cross term is important to understand  the entanglement between center symmetry and chiral symmetry, as pointed out by \cite{Fukushima04}.
In fact, the one-loop calculation leads to 
$C_1=0.97434  N_f>0$ and 
$F_1=- 0.106103 N_f<0$ ($\mu_q=0$ case) which appears to be a good indication for this purpose and serves as the initial condition in solving the flow equation.

In the paper by Schaefer, Pawlowski and Wambach \cite{SPW07}, 
a sort of back-reaction from quarks has been introduced to improve the effective potential of the Polyakov loop, while the NJL coupling remains local. 
In contrast, this paper introduces a back-reaction from gluons to improve the NJL interaction, leading to the nonlocal NJL coupling. 
However, this does not mean that two treatments are considered to be alternative. 
In the presence of dynamical quarks, the running coupling $\alpha$ is changed due to fermionic contributions.  In \cite{SPW07}, this effect has been taken into account as a modification of the expansion coefficient in the effective potential of the Polyakov loop, resulting in e.g., the $N_f$ flavor-dependent deconfinement temperature $T_d(N_f)$.
Remembering that the input of our analysis is just a running coupling, a sort of back-reaction from quarks considered in \cite{SPW07} is easily included into our framework by using the running coupling modified by quark contributions.  
Thus, the treatment in this paper is already able to take into account  back-reactions from quarks and gluons mentioned above.

This section is a sketch of our strategy of understanding the entanglement between center symmetry and chiral symmetry.
The detailed analysis will be given in a subsequent paper. 
\footnote{
It is known that appearance of a mixed-term $\sigma^2\tilde\varphi$ plays an essential role in the chiral-confinement
entanglement. Such a term appears in the original PNJL model and leads to the 2 crossovers happening almost simultaneously. 
In the following paper posted to the archive after this paper was submitted for publication, it has been shown that an effective Polyakov loop-dependent four-quark interaction derived by this paper yields stronger correlation between the chiral and deconfinement transitions, making $T_\chi \sim T_d$ more tightly, than the usual PNJL model. 
\\
Y. Sakai, T. Sasaki, H. Kouno and M. Yahiro,
Entanglement between deconfinement transition and chiral symmetry restoration, 
e-Print: arXiv:1006.3648 [hep-ph]. 
}

\section{Conclusion and discussion}\label{sec:conclusion}

In this paper, we have presented a reformulation of QCD and suggested a framework for deriving a low-energy effective theory of QCD which enables one to study the deconfinement/confinement and chiral-symmetry restoration/breaking crossover transition simultaneously on an equal footing.
A resulting low-energy effective theory based on this framework can be regarded as a modified (improved) version of the nonlocal PNJL model \cite{HRCW08}.  
In our framework, the basic ingredients are a reformulation of QCD based on new variables and the flow equation of the Wetterich type for the Wilsonian renormalization group.

A lesson we learned in this study is that a perturbative (one-loop) result can be a good initial condition for solving the flow equation of the renormalization group to obtain the non-perturbative result.  
In gluodynamics, recently, it has been demonstrated \cite{MP08,BGP07} that the existence of confinement transition, i.e., recovery of the center symmetry signaled by the vanishing Polyakov loop average can be shown by approaching the phase transition point from the high-temperature deconfinement phase in which the center symmetry is spontaneously broken.
Indeed, the effective potential for the Polyakov loop obtained in the one-loop calculation which we call the Weiss potential  leads to the non-vanishing Polyakov loop average, i.e., spontaneous breaking of the center symmetry. 
 

For gluon sector, to understand the existence of confinement transition by approaching from the deconfinement side, we have given the Landau-Ginzburg description in the neighborhood of the (crossover) phase transition point by analyzing the flow equation of the functional renormalization group. 
The deconfinement/confinement phase transition is consistent with the second order transition for $G=SU(2)$, while the first order transition is expected for $G=SU(3)$.
The detailed study of the  $SU(3)$ case will be given in a subsequent paper.

The input for solving the flow equation was just a running gauge coupling constant, in sharp contrast to the PNJL model including several parameters. 
From the viewpoint of a first-principle derivation, this is superior to phenomenological models with many input parameters.

For quark sector, it is possible  to obtain the chiral-symmetry breaking/restoration transition from the first principle.
However, we need more hard works, especially, to discuss the QCD phase diagram at finite density and the critical endpoint. 
A possibility in this direction from the first principle of QCD was demonstrated in one-flavor QCD based on the FRG \cite{Braun09}.  
It will be possible to treat chiral dynamics and confinement on an equal footing based on our framework along this line \cite{Kondo10}. 
Still, however, we must overcome some technical issues to achieve the goal of understanding full phase structures of QCD.
The detailed studies will be hopefully given in a subsequent paper.

{\em Acknowledgements --} 
The author would like to thank the Yukawa Institute for Theoretical Physics at Kyoto University where the YITP workshop  ``New Frontiers in QCD 2010  (NFQCD10)'' was held. 
He thanks the organizers and the participants, especially, Hideo Suganuma, Kenji Fukushima, Akira Ohnishi, Hiroshi Toki,  Wolfram Weise and Akihiro Shibata for discussions and comments on his talk, which were useful to complete this work. 
Thanks are due to Jan Pawlowski for discussions and correspondences on papers \cite{MP08,BGP07}, and Pengming Zhang for translating the original paper \cite{DG79} from Chinese into English. 
He is grateful to High Energy Physics Theory Group and Theoretical Hadron Physics Group in the University of Tokyo, especially, Prof. Tetsuo Hatsuda for kind hospitality extended to him on sabbatical leave between April 2009 and March 2010.
This work is financially supported by Grant-in-Aid for Scientific Research (C) 21540256 from Japan Society for the Promotion of Science
(JSPS).

\begin{appendix}
\section{Reformulation of QCD}\label{app:reformulation}

We apply the decomposition (\ref{decomp}) to QCD Lagrangian. 

The quark part is decomposed according to (\ref{decomp}) as 
\begin{align}
  \mathscr{L}_{q}  
:=&  \bar{\psi} (i\gamma^\mu \mathcal{D}_\mu[\mathscr{A}] -\hat{m}_0  + i \mu \gamma^0) \psi
\nonumber\\
=&   \bar{\psi} (i\gamma^\mu \mathcal{D}_\mu[\mathscr{V}] -\hat{m}_0 + i \mu \gamma^0) \psi +   g\bar{\psi} \gamma^\mu \mathscr{X}_\mu  \psi
 ,
\end{align}
where the covariant derivative $\mathcal{D}_\mu[\mathscr{V}]$ is defined by
\begin{equation}
 \mathcal{D}_\mu[\mathscr{V}] :=  \partial_\mu - ig \mathscr{V}_\mu   .
\end{equation}

The Yang-Mills part is treated as follows. 
For the general decomposition $\mathscr{A}_\mu(x)=\mathscr{V}_\mu(x)+\mathscr{X}_\mu(x)$, the field strength $\mathscr{F}_{\mu\nu}$ is decomposed as 
\begin{align}
  \mathscr{F}_{\mu\nu}[\mathscr{A}] :=& \partial_\mu \mathscr{A}_\nu - \partial_\nu \mathscr{A}_\mu -i g [ \mathscr{A}_\mu , \mathscr{A}_\nu ]
\nonumber\\
=& \mathscr{F}_{\mu\nu}[\mathscr{V}]  + \partial_\mu \mathscr{X}_\nu - \partial_\nu \mathscr{X}_\mu -i g [ \mathscr{V}_\mu , \mathscr{X}_\nu ] 
\nonumber\\&
-i g [ \mathscr{X}_\mu , \mathscr{V}_\nu ]
-i g [ \mathscr{X}_\mu , \mathscr{X}_\nu ]
\nonumber\\
=& \mathscr{F}_{\mu\nu}[\mathscr{V}]  + D_\mu[\mathscr{V}] \mathscr{X}_\nu - D_\nu[\mathscr{V}] \mathscr{X}_\mu 
\nonumber\\&
-i g [ \mathscr{X}_\mu , \mathscr{X}_\nu ] ,
\end{align}
where the covariant derivative $D_\mu[\mathscr{V}]$ in the background field $\mathscr{V}_\nu$ is defined by
\begin{align}
D_\mu[\mathscr{V}] := \partial_\mu \mathbf{1} -ig [ \mathscr{V}_\mu , \cdot ] ,
\end{align}
or, equivalently, 
\begin{equation}
  D_\mu[\mathscr{V}]^{AC} := \partial_\mu \delta^{AC} + g f^{ABC} \mathscr{V}_\mu^B .
\end{equation}
The Lagrangian density 
$\mathscr{L}_{YM} =  -\frac{1}{4} \mathscr{F}_{\mu\nu}[\mathscr{A}] \cdot \mathscr{F}^{\mu\nu}[\mathscr{A}]$ 
of the Yang-Mills theory is decomposed as
\begin{align}
   \mathscr{L}_{YM} 
=& -\frac{1}{4} \mathscr{F}_{\mu\nu}[\mathscr{A}]^2
\\
=& -\frac{1}{4} \mathscr{F}_{\mu\nu}[\mathscr{V}]^2
\nonumber\\&
- \frac{1}{2} \mathscr{F}^{\mu\nu}[\mathscr{V}] \cdot (D_\mu[\mathscr{V}] \mathscr{X}_\nu - D_\nu[\mathscr{V}] \mathscr{X}_\mu)
\nonumber\\&
- \frac{1}{4} (D_\mu[\mathscr{V}] \mathscr{X}_\nu - D_\nu[\mathscr{V}] \mathscr{X}_\mu)^2 
\nonumber\\&
+ \frac{1}{2}  \mathscr{F}_{\mu\nu}[\mathscr{V}] \cdot ig[ \mathscr{X}^\mu , \mathscr{X}^\nu ]
\nonumber\\&
+ \frac{1}{2}  (D_\mu[\mathscr{V}] \mathscr{X}_\nu - D_\nu[\mathscr{V}] \mathscr{X}_\mu) \cdot ig[ \mathscr{X}^\mu , \mathscr{X}^\nu ]
\nonumber\\&
- \frac14 (i g [ \mathscr{X}_\mu , \mathscr{X}_\nu ])^2
.
\end{align}
Here the third term on the right-hand side of the above equation is rewritten using integration by parts (or up to   total derivatives) as 
\begin{align}
 & \frac{1}{4} (D_\mu[\mathscr{V}] \mathscr{X}_\nu - D_\nu[\mathscr{V}] \mathscr{X}_\mu)^2
\nonumber\\
=&   \frac{1}{2} (- \mathscr{X}_\mu \cdot D_\nu[\mathscr{V}]D^\nu[\mathscr{V}] \mathscr{X}^\mu + \mathscr{X}_\mu  \cdot D_\nu[\mathscr{V}]D^\mu[\mathscr{V}]   \mathscr{X}^\nu )  
\nonumber\\
=& \frac{1}{2} \mathscr{X}^\mu  \cdot  \{ - D_\rho[\mathscr{V}]D^\rho[\mathscr{V}] g_{\mu\nu} +  D_\nu[\mathscr{V}]D_\mu[\mathscr{V}] \} \mathscr{X}^\nu
\nonumber\\
=& \frac{1}{2} \mathscr{X}^{\mu A}  \{ - (D_\rho[\mathscr{V}]D_\rho[\mathscr{V}])^{AB} g_{\mu\nu} 
- [ D_\mu[\mathscr{V}], D_\nu[\mathscr{V}]]^{AB}
\nonumber\\&
+  (D_\mu[\mathscr{V}]  D_\nu[\mathscr{V}])^{AB} \} \mathscr{X}^{\nu B} 
\nonumber\\
=& \frac{1}{2} \mathscr{X}^{\mu A}  \{ - (D_\rho[\mathscr{V}]D_\rho[\mathscr{V}])^{AB} g_{\mu\nu} 
+ gf^{ABC} \mathscr{F}_{\mu\nu}^{C}[\mathscr{V}] 
\nonumber\\&
+  D_\mu[\mathscr{V}]^{AC} D_\nu[\mathscr{V}]^{CB} \} \mathscr{X}^{\nu B}  ,
\end{align}
where we have used
\begin{align}
[ D_\mu[\mathscr{V}], D_\nu[\mathscr{V}]]^{AB}
=&  [ D_\mu[\mathscr{V}]^{AC}, D_\nu[\mathscr{V}]^{CB}] 
\nonumber\\ 
=& - gf^{ABC} \mathscr{F}_{\mu\nu}^{C}[\mathscr{V}]  .
\end{align}
Thus we obtain
\begin{align}
   \mathscr{L}_{YM} 
=& - \frac{1}{4} \mathscr{F}_{\mu\nu}[\mathscr{V40}]^2
\nonumber\\&
- \frac{1}{2} \mathscr{F}^{\mu\nu}[\mathscr{V}] \cdot (D_\mu[\mathscr{V}] \mathscr{X}_\nu - D_\nu[\mathscr{V}] \mathscr{X}_\mu)
\nonumber\\&
-     \frac{1}{2} \mathscr{X}^{\mu A}  W_{\mu\nu}^{AB} \mathscr{X}^{\nu B} 
\nonumber\\&
+ \frac{1}{2}  (D_\mu[\mathscr{V}] \mathscr{X}_\nu - D_\nu[\mathscr{V}] \mathscr{X}_\mu) \cdot ig[ \mathscr{X}^\mu , \mathscr{X}^\nu ]
\nonumber\\&
- \frac14 (i g [ \mathscr{X}_\mu , \mathscr{X}_\nu ])^2 
 ,
\end{align}
where we have defined
\begin{align}
W_{\mu\nu}^{AB}  :=& - (D_\rho[\mathscr{V}]D^\rho[\mathscr{V}])^{AB} g_{\mu\nu} 
+ 2gf^{ABC} \mathscr{F}_{\mu\nu}^{C}[\mathscr{V}] 
\nonumber\\&
+  D_\mu[\mathscr{V}]^{AC} D_\nu[\mathscr{V}]^{CB} . 
\label{W2}
\end{align}

In the usual background field method, the $\mathcal{O}(\mathscr{X})$ term is eliminated by requiring that the back ground field $\mathscr{V}$ satisfies the equation of motion $D_\mu[\mathscr{V}] \mathscr{F}^{\mu\nu}[\mathscr{V}]=0$:
\begin{align}
& \frac{1}{2} \mathscr{F}^{\mu\nu}[\mathscr{V}] \cdot (D_\mu[\mathscr{V}] \mathscr{X}_\nu - D_\nu[\mathscr{V}] \mathscr{X}_\mu)
\nonumber\\
=& - \frac{1}{2} ( D_\mu[\mathscr{V}] \mathscr{F}^{\mu\nu}[\mathscr{V}] \cdot  \mathscr{X}_\nu - D_\nu[\mathscr{V}] \mathscr{F}^{\mu\nu}[\mathscr{V}] \cdot  \mathscr{X}_\mu) = 0 .
\end{align}
In our framework, $\mathscr{V}$ do not necessarily satisfy the equation of motion. Nevertheless, the $\mathcal{O}(\mathscr{X})$ term vanishes from the defining equations which specify the decomposition.  For $G=SU(2)$,  $D_\mu[\mathscr{V}]\bm{n}=0$ and $\mathscr{X}_\mu \cdot \bm{n}=0$  lead to
\begin{align}
&   \mathscr{F}^{\mu\nu}[\mathscr{V}] \cdot (D_\mu[\mathscr{V}] \mathscr{X}_\nu)  
= G^{\mu\nu} \bm{n} \cdot (D_\mu[\mathscr{V}] \mathscr{X}_\nu)
\nonumber\\
=&    G^{\mu\nu} [\partial_\mu (\mathscr{X}_\nu \cdot \bm{n}) -  \mathscr{X}_\nu \cdot D_\mu[\mathscr{V}]\bm{n}] = 0
 .
\end{align}
In order for the reformulated theory written in terms of new variables to be equivalent to the original QCD, we must impose the reduction condition \cite{KMS06}:
\begin{equation}
  D_\mu[\mathscr{V}] \mathscr{X}^\mu = 0 .
  \label{reduction-cond}
\end{equation}
This eliminate the last term of $W_{\mu\nu}^{AB}$ in (\ref{W2}). 

Moreover, the $\mathcal{O}(\mathscr{X}^3)$ term is absent, i.e., 
\begin{equation}
\frac{1}{2}  (D_\mu[\mathscr{V}] \mathscr{X}_\nu - D_\nu[\mathscr{V}] \mathscr{X}_\mu) \cdot ig[ \mathscr{X}^\mu , \mathscr{X}^\nu ] = 0 
 ,
\end{equation}
since $D_\mu[\mathscr{V}] \mathscr{X}_\nu - D_\nu[\mathscr{V}] \mathscr{X}_\mu$ is orthogonal to $[ \mathscr{X}^\mu , \mathscr{X}^\nu ]$. See \cite{KMS06,KSM08,Kondo08}.

Thus, the Yang-Mills Lagrangian density reads
\begin{align}
   \mathscr{L}_{YM}  
=& -\frac{1}{4} \mathscr{F}_{\mu\nu}^A[\mathscr{V}]^2
-  \frac{1}{2} \mathscr{X}^{\mu A}  Q_{\mu\nu}^{AB} \mathscr{X}^{\nu B} 
\nonumber\\&
- \frac14 (i g [ \mathscr{X}_\mu , \mathscr{X}_\nu ])^2
 ,
\end{align}
where we have defined 
\begin{equation}
Q_{\mu\nu}^{AB}  := - (D_\rho[\mathscr{V}]D^\rho[\mathscr{V}])^{AB} g_{\mu\nu} 
+ 2gf^{ABC} \mathscr{F}_{\mu\nu}^{C}[\mathscr{V}] .
\end{equation}

For $G= SU(2)$,  the $\mathcal{O}(\mathscr{X}^3)$ term is absent, because $\mathscr{F}_{\mu\nu}[\mathscr{V}]$ and $-ig [ \mathscr{X}_\mu , \mathscr{X}_\nu]$ 
are   parallel to $\bm{n}$  (this is also the case for the sum
$\mathscr{F}_{\mu\nu}[\mathscr{V}] -i g [ \mathscr{X}_\mu , \mathscr{X}_\nu]$),
while $D_\mu[\mathscr{V}] \mathscr{X}_\nu - D_\nu[\mathscr{V}] \mathscr{X}_\mu$
 is orthogonal to $\bm{n}$ (which follows from the fact $\bm{n} \cdot \mathscr{X}_\mu=0$). 
 For $G=SU(2)$, therefore, we have
\begin{equation}
\mathscr{F}_{\mu\nu}^{C}[\mathscr{V}]=n^C  G_{\mu\nu}[\mathscr{V}] .
\end{equation}
Then the $SU(2)$ gluon part is rewritten into 
\begin{align}
   \mathscr{L}_{YM}  
=& -\frac{1}{4} (G_{\mu\nu}[\mathscr{V}])^2
-     \frac{1}{2} \mathscr{X}^{\mu A}  Q_{\mu\nu}^{AB}[\mathscr{V}] \mathscr{X}^{\nu B} 
\nonumber\\&
- \frac14 (i g [ \mathscr{X}_\mu , \mathscr{X}_\nu ])^2 
,
\end{align}
where
\begin{equation}
Q_{\mu\nu}^{AB}[\mathscr{V}]  = - (D_\rho[\mathscr{V}]D^\rho[\mathscr{V}])^{AB} g_{\mu\nu} 
+ 2g\epsilon^{ABC} n^C  G_{\mu\nu}[\mathscr{V}] .
\label{W}
\end{equation}

\section{Coefficients in the effective potential}\label{app:coefficient}

We expand $\hat V_{T,\hat k}$ defined by
\begin{align}
 \hat V_{T,\hat k}
 =& \hat V_{W} + 4  \int_{0}^{\hat k_{T}} \frac{d\hat p \hat p^2}{(2\pi)^2} \{ \ln (1- 2 e^{-\hat  k_{T}} \cos \varphi + e^{-2\hat k_{T}}) 
\nonumber\\&
- \ln (1- 2 e^{-\hat p} \cos \varphi + e^{-2\hat p}) \} 
  ,
\end{align}
in power series of $\tilde \varphi$ by using the expansion  
$
  - \cos \varphi = - \cos (\pi+\tilde \varphi) =  \cos ( \tilde \varphi)
  = 1 - \frac12 \tilde \varphi^2 + \frac{1}{24} \tilde \varphi^4 + O(\tilde \varphi^6)
$
as follows.
\begin{widetext}
\begin{align}
 \hat V_{T,\hat k}
 =& \hat V_{W} +   \int_{0}^{\hat k_{T}} \frac{d\hat p \hat p^2}{\pi^2} \Big\{ \ln \left[ 1+ 2 e^{-\hat  k_{T}}  - e^{-\hat  k_{T}} \tilde \varphi^2 + \frac{1}{12} e^{-\hat  k_{T}} \tilde \varphi^4  + e^{-2\hat k_{T}} + O(\tilde \varphi^6) \right]
  \nonumber\\
 & - \ln \left[ 1+ 2 e^{-\hat p}   - e^{-\hat p} \tilde \varphi^2 + \frac{1}{12} e^{-\hat p} \tilde \varphi^4   + e^{-2\hat p} + O(\tilde \varphi^6) \right] \Big\} 
 \nonumber\\
 =& \hat V_{W} +   \int_{0}^{\hat k_{T}} \frac{d\hat p \hat p^2}{\pi^2} \Big\{ \ln \left[ (1+  e^{-\hat  k_{T}} )^2 - e^{-\hat  k_{T}} \tilde \varphi^2 + \frac{1}{12} e^{-\hat  k_{T}} \tilde \varphi^4   + O(\tilde \varphi^6) \right]
  \nonumber\\
 & - \ln \left[ (1+  e^{-\hat p})^2   - e^{-\hat p} \tilde \varphi^2 + \frac{1}{12} e^{-\hat p} \tilde \varphi^4  + O(\tilde \varphi^6)   \right] \Big\} 
 \nonumber\\
 =& \hat V_{W} +   \int_{0}^{\hat k_{T}} \frac{d\hat p \hat p^2}{\pi^2} \Big\{ \ln  (1+  e^{-\hat  k_{T}} )^2 + \ln \left[ 1 - \frac{e^{-\hat  k_{T}}}{(1+  e^{-\hat  k_{T}} )^2} \tilde \varphi^2 + \frac{e^{-\hat  k_{T}}}{12(1+  e^{-\hat  k_{T}} )^2}  \tilde \varphi^4  + O(\tilde \varphi^6)  \right] 
  \nonumber\\
 & - \ln  (1+  e^{-\hat p})^2 - \ln \left[ 1  - \frac{e^{-\hat p}}{(1+  e^{-\hat p})^2} \tilde \varphi^2 + \frac{e^{-\hat p}}{12(1+  e^{-\hat p})^2}  \tilde \varphi^4   + O(\tilde \varphi^6) \right] \Big\} 
  .
\end{align}
By using 
$\log(1+x)=x-\frac12 x^2+O(x^3)$, therefore, $\hat V_{T,\hat k}$ has  the polynomial expansion:
\begin{align}
 \hat V_{T,\hat k}
 = A_{0,k} +  \frac{A_{2,k}}{2} \tilde \varphi^2 + \frac{A_{4,k}}{4!} \tilde \varphi^4 + O(\tilde \varphi^6)
  ,
\end{align}
where the coefficient is given by the integral form:
\begin{align}
\frac{A_{2,k}}{2}
=& - \frac16 +  \int_{0}^{\hat k_{T}} \frac{d\hat p \hat p^2}{\pi^2} 
\left[ \frac{e^{-\hat p}}{(1+  e^{-\hat p})^2} - \frac{e^{-\hat  k_{T}}}{(1+  e^{-\hat  k_{T}} )^2}  \right]
 ,
\nonumber\\
 \frac{A_{4,k}}{4!}  =& \frac{1}{12\pi^2} -  \int_{0}^{\hat k_{T}} \frac{d\hat p \hat p^2}{\pi^2} 
\left[ \frac{-6e^{-2\hat p}+e^{-\hat p}(1+  e^{-\hat p})^2}{12(1+  e^{-\hat p})^4} - \frac{-6e^{-2\hat  k_{T}}+e^{-\hat  k_{T}}(1+  e^{-\hat  k_{T}} )^2}{12(1+  e^{-\hat  k_{T}} )^4}  \right]
 ,
\nonumber\\
A_{0,k} =&     \int_{0}^{\hat k_{T}} \frac{d\hat p \hat p^2}{\pi^2} \Big\{ \ln  (1+  e^{-\hat  k_{T}} )^2  - \ln  (1+  e^{-\hat p})^2 \Big\}
 .
\end{align}
The integration can be performed analytically and the coefficient has the closed form:
\begin{align}
\frac{A_{2,k}}{2}
=&  - \frac16 +  \frac{1}{\pi^2}  \left[ -\frac{e^s s^3}{3 \left(1+e^s\right)^2}+\frac{e^s
   s^2}{1+e^s}-2 \log \left(1+e^s\right) s-2
   \text{Li}_2\left(-e^s\right)-\frac{\pi ^2}{6} \right] \Big|_{s=\hat k_{T}}
 ,
\nonumber\\
 \frac{A_{4,k}}{4!}  =& \frac{1}{12\pi^2} 
 + \frac{e^{2 s} \left(-2 s^3+\left(s^2+6\right) \cosh (s)
   s+6 s+3 \left(s^2-2\right) \sinh (s)-3 \sinh (2
   s)\right)}{18 \left(1+e^s\right)^4 \pi ^2} \Big|_{s=\hat k_{T}}
 ,
\end{align}
\end{widetext}
where ${\rm Li}_n(z)={\rm  PolyLog}[n,z]$ is the polylogarithm function, and in particular, the dilogarithm satisfies ${\rm Li}_2(z) = \int_{z}^{0} \frac{\log(1-t)}{t} dt$ which is known as the Spence integral.

Note that the function $\frac{e^{-\hat x}}{(1+  e^{-\hat x})^2}$ is monotonically decreasing in $x$ and hence the second term in $A_{2,k}$ is positive (non-negative). 
The coefficient $A_{2,k}$ is negative and monotonically increasing in $k$ and approaches zero for $k \rightarrow \infty$. 
\begin{equation}
\frac{A_{2,k}}{2} = -\frac16, \quad
-\frac16 \le \frac{A_{2,k}}{2} < 0 \quad {\rm for} \quad k \in [0, \infty) ,
\end{equation}
or
\begin{equation}
-\frac13 \le \frac{\partial^2}{\partial \tilde\varphi^2} \hat V_{T,\hat k} \Big|_{\tilde\varphi=0} < 0 \quad {\rm for} \quad k \in [0, \infty).
\end{equation}
This is because
\begin{align}
 \int_{0}^{\hat k_{T}}  d\hat p \hat p^2  
 \frac{e^{-\hat p}}{(1+  e^{-\hat p})^2} 
&\rightarrow  \int_{0}^{\infty}  d\hat p \hat p^2  
 \frac{e^{-\hat p}}{(1+  e^{-\hat p})^2} 
=  \frac16 \pi^2   
 ,
\\ 
 \int_{0}^{\hat k_{T}}  d\hat p \hat p^2  
 \frac{e^{-\hat  k_{T}}}{(1+  e^{-\hat  k_{T}} )^2}   
&= \frac13 \hat k_{T}^3  \frac{e^{-\hat  k_{T}}}{(1+  e^{-\hat  k_{T}} )^2}   \rightarrow 0
.
\end{align}

The coefficient $A_{4,k}$ is positive and approaches $0$ for $k \rightarrow \infty$, although $A_{4,k}$ is not monotonically decreasing in $k$. 
\begin{equation}
\frac{A_{4,0}}{4!}= \frac{1}{12\pi^2}  , \quad
\frac{A_{4,k}}{4!} > 0 \quad {\rm for} \quad k \in [0, \infty).
\end{equation}
This is because
\begin{align}
  & \int_{0}^{\hat k_{T}}  d\hat p \hat p^2  
\left[ \frac{-6e^{-2\hat p}+e^{-\hat p}(1+  e^{-\hat p})^2}{12(1+  e^{-\hat p})^4}    \right]
 \rightarrow    \frac{1}{12}   
 ,
\nonumber\\
 & \int_{0}^{\hat k_{T}}  d\hat p \hat p^2 
\left[ \frac{-6e^{-2\hat  k_{T}}+e^{-\hat  k_{T}}(1+  e^{-\hat  k_{T}} )^2}{12(1+  e^{-\hat  k_{T}} )^4}  \right]
\\&
= \frac13 \hat k_{T}^3  \left[ \frac{-6e^{-2\hat  k_{T}}+e^{-\hat  k_{T}}(1+  e^{-\hat  k_{T}} )^2}{12(1+  e^{-\hat  k_{T}} )^4}  \right] \rightarrow 0 
.
\end{align}

\section{Flow equation for the coefficient}\label{app:flow-equation}

Suppose that $\Delta \hat V_{\hat k}$ is of the form:
\begin{equation}
       \Delta \hat V_{\hat k}     
=    a_{0,k} + a_{1,k}  \tilde\varphi +  \frac{a_{2,k}}{2} \tilde \varphi^2  
+ \frac{a_{3,k}}{3!} \tilde \varphi^3 
+ \frac{a_{4,k}}{4!} \tilde \varphi^4 + O(\tilde \varphi^6)
 .
\end{equation}
The left-hand side of the flow equation reads
\begin{align}
     \partial_{\hat k}  \Delta \hat V_{\hat k}     
=&   \partial_{\hat k}  a_{0,k} + \partial_{\hat k} a_{1,k}  \tilde \varphi  +  \partial_{\hat k} \frac{a_{2,k}}{2} \tilde \varphi^2  
+ \partial_{\hat k} \frac{a_{3,k}}{3!} \tilde \varphi^3 
\nonumber\\&
+ \partial_{\hat k} \frac{a_{4,k}}{4!} \tilde \varphi^4 
+ O(\tilde \varphi^6)
 .
\end{align}
The flow equation $\partial_{\hat k} a_{n,k}$ for the coefficient  of $\tilde\varphi^n$ is extracted by differentiating both sides of the flow equation $n$ times and by putting $\tilde\varphi=0$.
The left-hand side is 
\begin{align}
\partial_{\hat k} a_{n,k} = \frac{\partial^n}{\partial \tilde\varphi^n} \partial_{\hat k}  \Delta \hat V_{\hat k}    \Big|_{\tilde\varphi=0} 
 .
\end{align}
Define  
\begin{equation}
  f(\varphi):= \frac{4\pi\alpha_k}{\hat k^2}   (\hat V_{T,\hat k}  + \Delta \hat V_{\hat k}  ) .
\end{equation}
The right-hand sides  of the flow equation $\partial_{\hat k} a_{n,k}$ are calculated from
\begin{align}
   \frac{\partial}{\partial \tilde\varphi} \left[  \frac{1}{1+\partial_\varphi^2 f(\varphi)} \right]  
=&  \frac{-\partial_\varphi^3 f(\varphi)}{[1+\partial_\varphi^2 f(\varphi)]^2}
 ,
\end{align}
\begin{align}
    \frac{\partial^2}{\partial \tilde\varphi^2} \left[  \frac{1}{1+\partial_\varphi^2 f(\varphi)} \right]  
=&  \frac{-\partial_\varphi^4 f(\varphi)}{[1+\partial_\varphi^2 f(\varphi)]^2} 
\nonumber\\&
-2  \frac{-[\partial_\varphi^3 f(\varphi)]^2}{[1+\partial_\varphi^2 f(\varphi)]^3}
 ,
\end{align}
\begin{align}
   \frac{\partial^3}{\partial \tilde\varphi^3} \left[  \frac{1}{1+\partial_\varphi^2 f(\varphi)} \right] 
=&  \frac{-\partial_\varphi^5 f(\varphi)}{[1+\partial_\varphi^2 f(\varphi)]^2} 
\nonumber\\&
-2  \frac{-3\partial_\varphi^4 f(\varphi) \partial_\varphi^3 f(\varphi)}{[1+\partial_\varphi^2 f(\varphi)]^3} 
\nonumber\\&
+ 6  \frac{-[\partial_\varphi^3 f(\varphi)]^3}{[1+\partial_\varphi^2 f(\varphi)]^4}
 ,
\end{align}
\begin{align}   
    \frac{\partial^4}{\partial \tilde\varphi^4} \left[  \frac{1}{1+\partial_\varphi^2 f(\varphi)} \right]  
=&  \frac{-\partial_\varphi^6 f(\varphi)}{[1+\partial_\varphi^2 f(\varphi)]^2} 
\nonumber\\&
-2  \frac{-4\partial_\varphi^5 f(\varphi) \partial_\varphi^3 f(\varphi)
-3[\partial_\varphi^4 f(\varphi)]^2}{[1+\partial_\varphi^2 f(\varphi)]^3} 
\nonumber\\&
+ 6  \frac{-6\partial_\varphi^4 f(\varphi)[\partial_\varphi^3 f(\varphi)]^2}{[1+\partial_\varphi^2 f(\varphi)]^4}
\nonumber\\&
-24   \frac{-  [\partial_\varphi^3 f(\varphi)]^4}{[1+\partial_\varphi^2 f(\varphi)]^5}
 , \cdots
\end{align}

If  $f$ is an even polynomial in $\tilde\varphi$, then the flow equation is simplified:
\begin{align}
\partial_{\hat k} a_{1,k} &\simeq 
\frac{\partial}{\partial \tilde\varphi} \left[  \frac{1}{1+\partial_\varphi^2 f(\varphi)} \right]   \Big|_{\tilde\varphi=0} 
=  0
 ,
\\     
\partial_{\hat k} a_{2,k} &\simeq  
\frac{\partial^2}{\partial \tilde\varphi^2} \left[  \frac{1}{1+\partial_\varphi^2 f(\varphi)} \right]   \Big|_{\tilde\varphi=0} 
\nonumber\\&
=   \frac{-\partial_\varphi^4 f(\varphi)}{[1+\partial_\varphi^2 f(\varphi)]^2}   \Big|_{\tilde\varphi=0} 
 ,
\\     
 \partial_{\hat k} a_{3,k} &\simeq
    \frac{\partial^3}{\partial \tilde\varphi^3} \left[  \frac{1}{1+\partial_\varphi^2 f(\varphi)} \right]  \Big|_{\tilde\varphi=0} 
=   0
 ,
\\     
\partial_{\hat k} a_{4,k} &\simeq
     \frac{\partial^4}{\partial \tilde\varphi^4} \left[  \frac{1}{1+\partial_\varphi^2 f(\varphi)} \right]   \Big|_{\tilde\varphi=0} 
\nonumber\\ 
&=   \frac{-\partial_\varphi^6 f(\varphi)}{[1+\partial_\varphi^2 f(\varphi)]^2}  \Big|_{\tilde\varphi=0} 
\nonumber\\&
+6  \frac{[\partial_\varphi^4 f(\varphi)]^2}{[1+\partial_\varphi^2 f(\varphi)]^3}  \Big|_{\tilde\varphi=0} 
 , ...
\end{align}
Therefore, with an initial condition, $a_{1,k} =0=a_{3,k}$ at $k=\Lambda$, the flow equations in the above
\begin{equation}
\partial_{\hat k} a_{1,k} =0, \quad \partial_{\hat k} a_{3,k}=0 .
\end{equation}
guarantee the solution  
\begin{equation}
  a_{1,k} \equiv 0, \quad   a_{3,k} \equiv 0 \quad (0 \le k \le \Lambda) .
\end{equation}

\end{appendix}
 


\begin{thebibliography}{99}

\bibitem{KS04}
J.B. Kogut and M.A. Stephanov,   
{\em The phases of quantum chromodynamics: From confinement to extreme environments} 
(Cambridge Univ. press, Cambridge, 2004).

\bibitem{YHM05}   
 K. Yagi, T. Hatsuda and Y. Miake,
{\em Quark-Gluon Plasma} 
(Cambridge Univ. press, Cambridge, 2005).


\bibitem{Polyakov78}
A.M. Polyakov, 
Phys. Lett. B {\bf 72}, 477--480 (1978).


\bibitem{Karsch02}
F. Karsch,
Lattice QCD at high temperature and density,
[hep-lat/0106019],
Lect. Notes Phys. {\bf 583}, 209--249 (2002).


\bibitem{KL94}
F. Karsch and E. Laermann,
Phys. Rev. D {\bf 50}, 6954 (1994).
\\
S. Aoki, M. Fukugita, S. Hashimoto, N. Ishizuka, Y. Iwasaki, K. Kanaya, Y. Kuramashi, H. Mino, M. Okawa, A. Ukawa and T. Yoshie, 
Phys. Rev. D{\bf 57}, 3910 (1998).


\bibitem{KL98}
F. Karsch and M. Lutgemeier, 
[hep-lat/9812023],
Nucl.Phys. B{\bf 550}, 449--464 (1999). 


\bibitem{Casher79}
A. Casher,
Phys. Lett. B {\bf 83}, 395 (1979).
\\
G. 't Hooft, 
in : Recent Developments in Gauge Theories,
eds. G. 't Hooft et al.
(Plenum Press, New York, 1980).


\bibitem{KST99}
J.B. Kogut,  M. A. Stephanov and D. Toublan,  
[hep-ph/9906346], 
Phys.Lett. B{\bf 464}, 183--191 (1999). 
\\
J.B. Kogut,  M.A. Stephanov, D. Toublan, J.J.M. Verbaarschot and  A. Zhitnitsky,  
[hep-ph/0001171], 
Nucl.Phys. B{\bf 582}, 477--513 (2000). 


\bibitem{MP07}
L. McLerran and R.D. Pisarski, 
arXiv:0706.2191 [hep-ph], 
Nucl.Phys. A{\bf 796}, 83--100 (2007). 


\bibitem{HKS10}
S. Hands, S. Kim and J.-I. Skullerud,  
arXiv:1001.1682 [hep-lat]. 


\bibitem{Lee72}
B.W. Lee, 
Chiral Dynamics
(Gordon \& Breach, New York, 1972).


\bibitem{NJL61}
Y. Nambu and G. Jona-Lasinio,
Phys. Rev. {\bf 122}, 345 (1961).
\\
Y. Nambu and G. Jona-Lasinio,
Phys. Rev. {\bf 124}, 246 (1961).


\bibitem{Klevansky92}
S.P. Klevansky,
Rev. Mod. Phys. {\bf 64}, 649 (1992).


\bibitem{HK94}
T. Hatsuda and T. Kunihiro, 
[hep-ph/9401310],  
Phys. Rept. {\bf 247}, 221--367 (1994). 


\bibitem{SV93}
E. Shuryak and J.J.M. Verbaarschot,
Nucl. Phys. A {\bf 560}, 306 (1993).


\bibitem{GL84}
J. Gasser and H. Leutwyler,
Ann. Phys. {\bf 158}, 142 (1984).


\bibitem{MO95}
P.N. Meisinger and M.C. Ogilvie, 
[hep-lat/9512011], 
Phys.Lett. B{\bf 379}, 163--168 (1996). 


\bibitem{MST03}
A. Mocsy, F. Sannino and K. Tuominen,  
[hep-ph/0308135], 
Phys. Rev. Lett.{\bf 92}, 182302 (2004). 
\\
A. Mocsy, F. Sannino and K. Tuominen,
e-Print: hep-ph/0401149.
 

\bibitem{Fukushima04}
K. Fukushima, 
[hep-ph/0310121],
Phys. Lett. B{\bf 591}, 277--284 (2004). 


\bibitem{MRAS06}
E. Megias, E. Ruiz Arriola, and L.L. Salcedo, 
[hep-ph/0412308],
Phys. Rev. D{\bf 74}, 065005 (2006). 
\\
E. Megias, E. Ruiz Arriola, and L.L. Salcedo,   
[hep-ph/0607338], 
Phys. Rev. D{\bf 74}, 114014 (2006). 
\\
E. Megias, E. Ruiz Arriola, and L.L. Salcedo, 
[hep-ph/0505215],
JHEP 0601,073 (2006). 



\bibitem{RTW05}
C. Ratti, M.A. Thaler and W. Weise,
[hep-ph/0506234],
Phys. Rev. D{\bf 73}, 014019 (2006). 


\bibitem{HRCW08}
T. Hell, S. R\"ossner, M. Cristoforetti and W. Weise, 
arXiv:0810.1099, 
Phys. Rev. D{\bf 79}, 014022 (2009).


\bibitem{SFR06}
C. Sasaki, B. Friman and K. Redlich, 
[hep-ph/0611147],
Phys. Rev. D{\bf 75}, 074013 (2007). 


\bibitem{BBRV07}
D. Blaschke, M. Buballa, A.E. Radzhabov and M.K. Volkov,
arXiv:0705.0384 [hep-ph], 
Yad.Fiz. {\bf 71}, 2012--2018 (2008), Phys. Atom. Nucl.{\bf 71}, 1981--1987 (2008). 
\\
D. Gomez Dumm, D.B. Blaschke, A.G. Grunfeld, and N.N. Scoccola,    
[hep-ph/0512218], 
Phys. Rev. D{\bf 73}, 114019 (2006). 


\bibitem{KKMY07}
K. Kashiwa, H. Kouno, M. Matsuzaki and M. Yahiro,    
arXiv:0710.2180 [hep-ph], 
Phys.Lett.B {\bf 662}, 26--32 (2008). 


\bibitem{SPW07}
B.-J. Schaefer,  J.M. Pawlowski and J. Wambach,  
arXiv:0704.3234 [hep-ph], 
Phys.Rev.D{\bf 76}, 074023 (2007). 


\bibitem{Wetterich93}
C. Wetterich, 
Phys. Lett. B{\bf 301}, 90--94 (1993).


\bibitem{BTW00}
J. Berges, N. Tetradis and C. Wetterich,  
[hep-ph/0005122],  
Phys. Rept. {\bf 363}, 223--386 (2002). 


\bibitem{Gies06}
H. Gies, 
hep-ph/0611146. 


\bibitem{Weiss81}
N. Weiss,
Phys. Rev. D{\bf 24}, 475--480 (1981).


\bibitem{MP08}
F. Marhauser and J.M. Pawlowski,
arXiv:0812.1144[hep-ph].


\bibitem{BGP07}
J. Braun, H. Gies and J.M. Pawlowski,
arXiv:0708.2413[hep-th], 
Phys. Lett. B{\bf 684}, 262--267 (2010)


\bibitem{BHMP09}
J. Braun, L.M. Haas, F. Marhauser and J.M. Pawlowski,
e-Print: arXiv:0908.0008 [hep-ph]. 


\bibitem{Cho80}
  Y.M. Cho,
Phys. Rev. D {\bf 21}, 1080--1088 (1980).
Y.M. Cho,
Phys. Rev. D {\bf 23}, 2415--2426 (1981). 


\bibitem{DG79}
  Y.S. Duan and M.L. Ge, 
Sinica Sci., {\bf 11}, 1072--1081 (1979). 


\bibitem{FN99} 
  L. Faddeev and A.J. Niemi,
[hep-th/9807069],
Phys. Rev. Lett. {\bf 82}, 1624--1627 (1999).


\bibitem{Shabanov99}
  S.V. Shabanov,
[hep-th/9903223],
Phys. Lett. B {\bf 458}, 322--330 (1999).
\\
  S.V. Shabanov, 
[hep-th/9907182],
Phys. Lett. B {\bf 463}, 263--272 (1999).


\bibitem{KMS06}
  K.-I. Kondo, T. Murakami and T. Shinohara,
[hep-th/0504107], 
Prog. Theor. Phys. {\bf 115}, 201--216 (2006). 


\bibitem{KMS05}
  K.-I. Kondo, T. Murakami and T. Shinohara,
[hep-th/0504198],
Eur. Phys. J. C {\bf 42}, 475--481 (2005).


\bibitem{Kondo06}
K.-I. Kondo,
[hep-th/0609166], 
Phys. Rev. D {\bf 74}, 125003 (2006). 

 
\bibitem{KSM08}
K.-I. Kondo, T. Shinohara and T. Murakami,
e-Print: arXiv:0803.0176 [hep-th],
Prog. Theor. Phys. {\bf 120},  1--50 (2008).


\bibitem{KKMSS05}
S. Kato,  K.-I. Kondo, T. Murakami,  A. Shibata and T. Shinohara,
e-Print: hep-ph/0504054.


\bibitem{KKMSSI05}
  S. Kato, K.-I. Kondo, T. Murakami, A. Shibata, T. Shinohara and S. Ito,
[hep-lat/0509069], 
Phys. Lett. B {\bf 632}, 326--332
 (2006).


\bibitem{IKKMSS06}
  S. Ito, S. Kato, K.-I. Kondo, T. Murakami, A. Shibata and T. Shinohara,  
[hep-lat/0604016], 
Phys. Lett. B {\bf 645}, 67--74  (2007).  
 

\bibitem{SKKMSI07}
A. Shibata, S. Kato, K.-I. Kondo, T. Murakami, T. Shinohara and  S. Ito,
arXiv:0706.2529 [hep-lat],
Phys.Lett. B{\bf 653}, 101--108 (2007). 


\bibitem{SKKMSI07b}
A. Shibata, S. Kato, K.-I. Kondo, T. Murakami, T. Shinohara, and S. Ito,
e-Print: arXiv:0710.3221 [hep-lat], 
POS(LATTICE-2007) 331.
\\
A. Shibata,  K.-I. Kondo, S. Kato, S. Ito, T. Shinohara, T. Murakami, 
arXiv:0810.0956 [hep-lat],
PoS(LATTICE 2008)268. 


\bibitem{SKKISF09}
A. Shibata, K.-I. Kondo, S. Kato, S. Ito,  T. Shinohara and N. Fukui,  
Talk given at 27th International Symposium on Lattice Field Theory (Lattice 2009), Beijing, China, 25-31 Jul 2009. 
arXiv:0911.4533 [hep-lat] 


\bibitem{KSSMKI08}
K.-I. Kondo, A. Shibata, T. Shinohara, T. Murakami, S. Kato and  S. Ito,
arXiv:0803.2451[hep-lat],
PLB{\bf 669}, 107--118 (2008).


\bibitem{SKS09}
A. Shibata, K.-I. Kondo and T. Shinohara,
arXiv:0911.5294 [hep-lat].


\bibitem{Cho00}
Y.M. Cho,
Phys. Rev. D {\bf 62}, 074009 (2000).


\bibitem{Kondo08}
  K.-I. Kondo,
arXiv:0801.1274 [hep-th],
Phys. Rev. D {\bf 77}, 085029 (2008).


\bibitem{KS08}
 K.-I. Kondo and A. Shibata, 
arXiv:0801.4203 [hep-th].


\bibitem{Kondo08b}
 K.-I. Kondo, 
arXiv:0802.3829 [hep-th],
J. Phys. G: Nucl. Part. Phys. {\bf 35}, 085001  (2008).


\bibitem{MO97}
 P.N. Meisinger and M.C. Ogilvie,
[hep-lat/9703009],
Phys.Lett. B407, 297--302 (1997). 
\\
 M. Engelhardt and H. Reinhardt,
[hep-th/9709115],
Phys.Lett. B430, 161--167 (1998). 
\\
H. Gies, 
[hep-th/0005252],
Phys. Rev. D63, 025013 (2001).
\\
J. Braun, H. Gies and H.-J. Pirner,
hep-ph/0610341.


\bibitem{Kondo01}
K.-I. Kondo,
[hep-th/0105299],
Phys. Lett. B{\bf 514}, 335--345 (2001). 
\\
K.-I. Kondo,
[hep-th/0306195], 
Phys. Lett. B{\bf 572}, 210--215 (2003). 
\\
L.D. Faddeev and A.J. Niemi,
[hep-th/0608111],
Nucl. Phys. B{\bf 776}, 38--65 (2007). 


\bibitem{dualsuper}
  Y. Nambu,
  Phys. Rev. D {\bf 10}, 4262--4268
 (1974).
\\
G. 't Hooft,
  in: High Energy Physics, edited by A. Zichichi 
(Editorice Compositori, Bologna, 1975).
\\
S. Mandelstam,
 Phys. Report  {\bf 23}, 245--249
 (1976).
\\
A.M. Polyakov,
  Nucl. Phys. B {\bf 120}, 429--458
 (1977).


\bibitem{EI82}
  Z.F. Ezawa and A. Iwazaki,
  Phys. Rev. D {\bf 25}, 2681--2689 (1982).


\bibitem{SY90} 
  T. Suzuki and I. Yotsuyanagi,
  Phys. Rev. D {\bf 42}, 4257--4260 (1990).


\bibitem{SNW94}
J.D. Stack, S.D. Neiman and R. Wensley,
[hep-lat/9404014],
Phys. Rev. D{\bf 50}, 3399--3405 (1994).
H.~Shiba and T.~Suzuki,
Phys.Lett.B{\bf 333}, 461--466  (1994).
  
  
\bibitem{AS99}
  K. Amemiya and H. Suganuma,
[hep-lat/9811035],
Phys. Rev. D{\bf 60}, 114509 (1999).
\\
  V.G. Bornyakov, M.N. Chernodub, F.V. Gubarev, S.M. Morozov and M.I. Polikarpov, 
[hep-lat/0302002],
 Phys. Lett. B{\bf 559}, 214--222 (2003).
 

\bibitem{SAIIMT02} 
H. Suganuma, K. Amemiya, H. Ichie, N. Ishii, H. Matsufuru and T.T. Takahashi,
[hep-lat/0407016], 
Nucl. Phys. B (Proc. Suppl.) {\bf 106}, 679--681 (2002)


\bibitem{AS00}
K. Amemiya and H. Suganuma,
[hep-lat/9909096], 
Nucl. Phys. B(Proc. Suppl.) {\bf 83}, 419--421 (2000).


\bibitem{KSSK10}
K.-I. Kondo, A. Shibata, T. Shinohara and S. Kato,
arXiv:1007.2696 [hep-th],
\\
A. Shibata  et al, in preparation. 


\bibitem{Miransky85}
V. A. Miransky, 
Nuovo Cim. A {\bf 90} (1985), 149--170.
\\
W. A. Bardeen, C. N. Leung and S. T. Love, 
Phys. Rev. Lett. {\bf 56}, 1230--1233 (1986).
\\
C. N. Leung, S.T. Love and W.A. Bardeen,  
Nucl. Phys. B273, 649--662 (1986). 


\bibitem{KMY89}
K.-I. Kondo, H. Mino, K. Yamawaki, 
Phys. Rev. D{\bf 39}, 2430--2433 (1989).
\\
T. Appelquist, M. Soldate, T. Takeuchi and L. C. R. Wijewardhana, in Proc. Johns Hopkins
Workshop on Current Problems in Particle Theory 12, Baltimore, June 8-10, 1988, ed. G.
Domokos and S. Kovesi-Domokos (World Scientific Pub. Co., Singapore, 1988), p. 197.


\bibitem{KKM89}
K.-I. Kondo,
Int. J. Mod. Phys. A{\bf 6}, 5447--5466 (1991).
\\
K.-I. Kondo,
Nucl. Phys. B{\bf 351}, 259--276 (1991).
\\
K.-I. Kondo and H. Nakatani, 
Nucl. Phys. B{\bf 351}, 236--258 (1991). 
\\
K.-I. Kondo, Y. Kikukawa, H. Mino
Phys. Lett. B220, 270--275 (1989). 


\bibitem{Kondo91}
K.-I. Kondo, 
Lecture given at Nagoya Spring School on Dynamical Symmetry Breaking, Nakatsugawa, Japan, Apr 23-27, 1991, 
CHIBA-EP-52. 


\bibitem{KSY91}
K.-I. Kondo,  S. Shuto, K. Yamawaki, 
Mod. Phys. Lett. A{\bf 6}, 3385--3396 (1991).


\bibitem{Aoki-etal99}
K-I. Aoki, K. Morikawa, J. Sumi, H. Terao and M. Tomoyose, 
e-Print: hep-th/9908042, 
Prog. Theor. Phys. {\bf 102}, 1151--1162 (1999). 


\bibitem{Litim00}
D.F. Litim,
[hep-th/0103195], 
Phys.Rev. D{\bf 64}, 105007 (2001). 
\\
D.F. Litim,
[hep-th/0005245], 
Phys.Lett. B{\bf 486},  92--99 (2000). 


\bibitem{BG05}
J. Braun and H. Gies,
[hep-ph/0512085],
Phys. Lett. B {\bf 645}, 53 (2007). 
J. Braun and H. Gies,
[hep-ph/0602226],
J. High Energy Phys. {\bf 06}, 024 (2006). 


\bibitem{FA02}
C.S. Fischer and R. Alkofer,
[hep-ph/0202202],
Phys. Lett. B{\bf 536}, 177--184 (2002). 


\bibitem{FMP08}
C. S. Fischer,  A. Maas and  J. M. Pawlowski,  
arXiv: 0810.1987 [hep-ph], 
Annals Phys. {\bf 324}, 2408-2437 (2009). 


\bibitem{Kondo09}
K.-I. Kondo, 
arXiv:0904.4897 [hep-th], 
Phys.Lett. B{\bf 678}, 322-330 (2009).
\\
K.-I. Kondo, 
arXiv:0907.3249 [hep-th], 
Prog. Theor. Phys. {\bf 122}, 1455--1475 (2009).
\\
K.-I. Kondo, 
arXiv:0909.4866 [hep-th].


\bibitem{Fierz}
Y. Takahashi,
Soryushiron Kenkyu, {\bf 66}, 127--139 (1982).

\bibitem{Gies01}
H. Gies, 
[hep-th/0102026], 
Phys. Rev. D{\bf 63}, 125023 (2001). 


\bibitem{Braun09}
J. Braun, 
arXiv:0810.1727 [hep-ph],
Eur.Phys.J.C. {\bf 64}, 459--482 (2009). 


\bibitem{Kondo10}
K.-I. Kondo,
Talk given at the YITP workshop  ``New Frontiers in QCD 2010  (NFQCD10)'' held at the Yukawa Institute for Theoretical Physics, Kyoto University, 4th March 2010. 




\end{thebibliography}
\end{document}